\documentclass{article}

\usepackage{psfrag,latexsym,amssymb,amsmath,natbib,color,epstopdf,lineno}
\usepackage{graphicx}%[dvips]

\usepackage{amsthm}
\newcommand{\bq}{\mbox{\boldmath $q$}} 
 
\newcommand{\bv}{\mbox{\boldmath $v$}} 
\newcommand{\nbQ}{\mbox{$\mathsf{Q}$}} 
\newcommand{\bQ}{\mbox{\boldmath $\mathsf{Q}$}}

\newcommand{\bS}{\mbox{\boldmath $\mathsf{S}$}} 
\newcommand{\nbP}{\mbox{$\mathsf{P}$}} 
\newcommand{\bP}{\mbox{\boldmath $\mathsf{P}$}} 
 
\newcommand{\bhP}{\mbox{\boldmath $\mathsf{\widehat P}$}} 
 
\newcommand{\ha}{{\widehat a}}

\newcommand{\hu}{{\widehat u}} 
\def\phivec{\mbox{\boldmath $\phi$}}

\def\uvec{\mbox{\boldmath $u$}}

\def\dd{{\, \rm{d}}}
\def\dr{{\rm{d}}}

\def\bra{\langle}
\def\ket{\rangle}
\def\p{\partial}

\def\beq{\begin{equation}}
\def\eeq{\end{equation}}
\def\la{\label}
\def\ii{{\rm i}}
\def\ee{{\rm e}}

\def\r#1{(\ref{#1})}
\newtheoremstyle{mio}{5pt}{5pt}{\itshape}{0pt}{\bfseries}{:}{.3em}{}%
\theoremstyle{mio}
\def\utau{u_\tau}

\def\ta{\widetilde{a}}
\def\tu{\widetilde{u}}
\def\tv{\widetilde{v}}

\def\figpath{./}
\def\spacce#1{\hskip #1pt}
\def\drawline#1#2{\raise 2.5pt\vbox{\hrule width #1pt height #2pt}}
\def\solid{\drawline{24}{.5}\nobreak}

\def\bdash{\hbox{\drawline{5}{.5}\spacce{2}}}

\def\dashed{\bdash\bdash\bdash\nobreak}

\def\trian{\raise 1.25pt\hbox{$\scriptstyle\triangle$}\nobreak}

\def\dtrian{\raise 1.25pt\hbox%
{$\scriptscriptstyle\bigtriangledown$}\nobreak}
\def\rtrian{\raise 1.25pt\hbox%
{$\scriptscriptstyle\vartriangleright$}\nobreak}

\def\squar{\raise 1.25pt\hbox{$\scriptstyle\Box$}\nobreak}

\def\diamon{\raise 1.25pt\hbox{$\scriptstyle\diamond$}\nobreak}

\newcommand{\soliddtrian}{$\blacktriangledown$\nobreak}

\def\linedtri1{\hbox{\bdash\hspace{-1.6mm}$\bigtriangleup$\hspace{-0.8mm}\bdash}\nobreak}
\def\soliddtrian1{$\blacktriangledown$\nobreak}
\def\solidrtrian2{$\blacktriangleright$\nobreak}
\def\solidltrian3{$\blacktriangleleft$\nobreak}

\title{\bf A Perron--Frobenius analysis of wall-bounded turbulence}

\author{Javier Jim\'enez\\
School of Aeronautics, U. Polit\'ecnica Madrid, 28040 Madrid Spain}
\date{\today}

\begin{document}
\maketitle
\begin{abstract}
The Perron--Frobenius operator (PFO) is adapted from dynamical-system theory to the study of
turbulent channel flow. It is shown that, as long as the analysis is restricted to the
system attractor, the PFO can be used to differentiate causality and coherence from simple
correlation without performing interventional experiments, and that the key difficulty
remains collecting enough data to populate the operator matrix. This is alleviated by
limiting the analysis to two-dimensional projections of the phase space, and developing a
series of indicators to choose the best parameter pairs from a large number of
possibilities. The techniques thus developed are applied to the study of bursting in the
inertial layer of the channel, with emphasis on the process by which bursts are reinitiated
after they have decayed. Conditional averaging over phase-space trajectories suggested by
the PFO shows, somewhat counter-intuitively, that a key ingredient for the burst recovery is
the development of a low-shear region near the wall, overlaid by a lifted shear layer. This is
confirmed by a computational experiment in which the control of the mean velocity profile by
the turbulence fluctuations is artificially relaxed. The behaviour of the mean velocity
profile is thus modified, but the association of low wall shear with the initiation of the
bursts is maintained.
\end{abstract}

% ----------------------------------------
%\keywords{wall turbulence, non-interventional causality, bursting.}

%\linenumbers
% --------------------------------------------------------------------------
\section{Introduction}\la{sec:intro}

\begin{figure}
%\vspace*{8mm}%
\centering
%
%\raisebox{0mm}{\includegraphics[height=.30\textwidth,clip]%
%{\figpath onesnap.pdf}}%
%\mylab{-.35\textwidth}{.25\textwidth}{(a)}%
%\hspace*{2mm}%
%%
%\raisebox{0mm}{\includegraphics[height=.32\textwidth,clip]%
%{\figpath twosnapschaos.pdf}}%
%\mylab{-.35\textwidth}{.25\textwidth}{(b)}%
\includegraphics[width=1\textwidth,clip]{\figpath 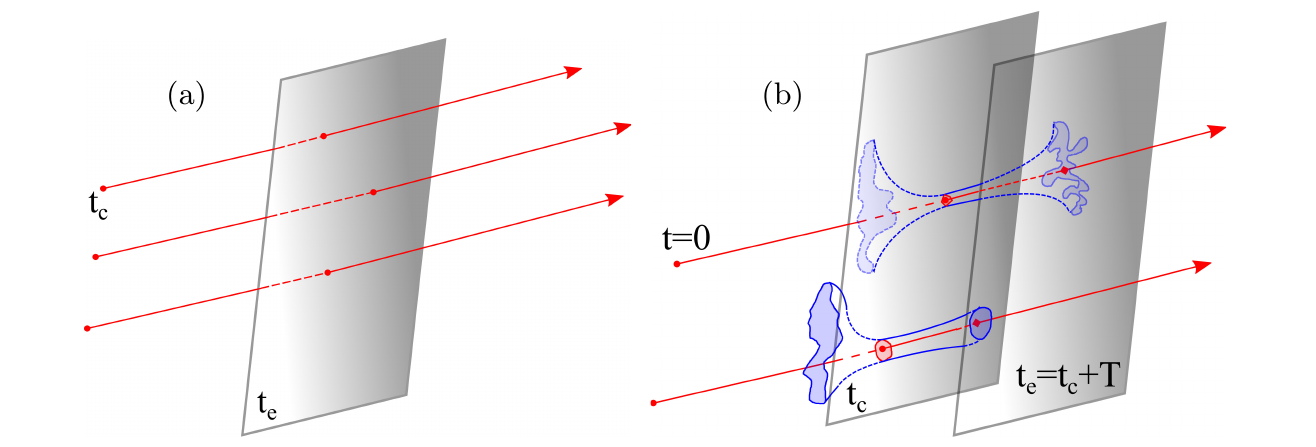}
\caption{%
(a) The deterministic view of a dynamical system. The shaded plane represents the full phase
space at one instant in time, and each trajectory is one possible evolution of the system.
(b) The effect of dissipation and chaos. Phase points have to be substituted by neighbourhoods, and
the flow becomes ill-conditioned both forward and backwards in time.
}
\label{fig:dynam_sys}
\end{figure}

There is widespread agreement that physical phenomena have causes, but less
consensus on what this may mean. Several questions come to mind. The first is whether
the concept of cause has any meaning when the equations of motion are known, and whether,
even if a definition could be agreed upon, it would be of any practical value. For example,
\cite{russ:12} argued that, if the temporal evolution of a dynamical system is described by
a set of deterministic differential equations, causality is equivalent to knowledge of the
initial conditions. This point of view can be traced to Newton and even to the classical
world, and implies that the only causes of the state of the system at time $t$ are the state of the
system at any previous time. This is sketched in figure \ref{fig:dynam_sys}(a). Disregarding
isolated singularities, any point $\bv(t_e)$ in phase space is the `effect' of all the
points $\bv(t_c<t_e)$ in a unique incoming trajectory. Conversely, $\bv(t_e)$ is the `cause'
of all the points in that trajectory for which $t>t_e$.

However, \cite{russ:12} was probably thinking about reversible Hamiltonian mechanics and,
although true in theory, his conclusions are not necessarily useful in more general cases.
Many mechanical systems are dissipative, and identifying the \cite{russ:12} cause of a
particular state implies integrating ill-posed equations backwards in time. This certainly
applies to Navier--Stokes turbulence, which is the system that mostly interests us here. Similarly,
\cite{russ:12} knew little about deterministic chaos, but we now understand that most dynamical
systems with many degrees of freedom are chaotic, and cannot in practice be uniquely
integrated forward. The evolution of turbulence is closer to figure \ref{fig:dynam_sys}(b),
in which $\bv(t_e)$ has been substituted by a small neighbourhood, and the forward and
backward trajectories become irregular or fractal cones formed by bundles of trajectories
that contain the causes and effects of the points in the neighbourhood of $\bv(t_e)$.
\citeauthor{russ:12}'s question can be recast as whether, in such situations, something is
retained of the deterministic picture in figure \ref{fig:dynam_sys}(a).

% ===========================================================
% ========================================================

A related problem is whether something can be said about causality without performing
interventional experiments. The usual answer is that it can not, because the correlations
that result from observations do not imply causation \citep{granger:69,Pearl:09}. But the
discussion in the previous paragraph suggests that this may not be the whole story, and that
a sufficiently careful observation of the temporal evolution of a system may lead to the
identification of the `causal' trajectories that cross a neighbourhood of interest
\citep{AngEtal:96}.

The coarse-graining inherent in figure \ref{fig:dynam_sys}(b) suggests that the dynamical
system can be simplified by partitioning the phase space into disjoint neighbourhoods of
finite size, at least for a fixed temporal horizon. This is common practice in chaotic
dynamical systems \citep{BeckSchl93}, and, although the reasons given are often that it avoids
singular measures in the statistics, and that numerical experiments are anyway discrete,
figure \ref{fig:dynam_sys}(b) suggests that there is a more fundamental justification. If
chaos prevents us from predicting the behaviour of infinitesimally close neighbouring
points, it makes little sense to insist on treating them as if they were different, and we
may as well consider finite-size neighbourhoods as our fundamental dynamical units. This has
important consequences for the definition of the system. The main one is that the system
propagator is substituted by the `symbolic' dynamics of how often and in which order the
system visits the different cells, and that the deterministic equations are substituted by
the transition probabilities incorporated in the Perron-Frobenius (transfer) operator introduced
in \S\ref{sec:PF}.

Turbulence is well suited for these techniques, because it is a chaotic deterministic system with
many degrees of freedom for which the (Navier--Stokes) equations are known. The
difficulty is not how to integrate the equations, which are in principle within reach of a
sufficiently large computer, but how to explain and predict turbulent flows in terms of
simpler rules. Direct simulations are exact but expensive, and we would like to have reduced
models that reproduce the flow, if not in full detail, at least well enough to provide
general rules about its future behaviour and, ideally, about how that behaviour could be
influenced. Shear-driven turbulence is particularly appropriate because it can be made
statistically steady, as in pipes or channels, but also because it is believed to be
partially controlled by linear processes \citep{ala:jim:06,McKSha2010,jim13_lin}, and at
least in part describable in terms of coherent structures that play the role of objects in a
dynamical system \citep{adr07,jim18}.

Especially interesting is the regeneration cycle of wall-bounded turbulence, whose
persistence has been explained by the interaction between the perturbations of the
streamwise and cross-flow velocities \citep{jim94,Hamilton95,Waleffe97}. There is fairly
general consensus that the wall-normal velocity generates fluctuations of the streamwise
velocity by deforming the mean shear, and that the shear interacts with the cross-flow
fluctuations to amplify them \citep{orr07a,jim13_lin,jimenez:2015}. But this amplification
is transient in most models \citep{but:far:93,far_ioa96,Schoppa02,jim13_lin}, and the
details of how the cycle closes after the burst decays are unclear. The elucidation of this
regeneration process is the underlying `application' of our investigation, but much of the
paper is dedicated to the development of the analysis procedure itself.

Note that the results of our analysis will not be causal in the sense of \cite{granger:69}
or \cite{Pearl:09}, since they involve no intervention from the observer. But we are more
interested in predictability and perhaps in coherence, and in the search for states of the
system that best allow us to draw conclusions from partial flow information. The fundamental
question of causality will be outsourced here to the equations of motion, and its direction
to the direction of time. The main purpose of our analysis is to identify flow
configurations in which the equations of motion give us the best possible information about
the future of the system without necessarily solving them in detail, and which could perhaps
lead to effective control strategies.

The organisation of the paper is as follows. Section 2 introduces the Perron-Frobenius
operator, which is particularised to a small-box turbulent channel in \S \ref{sec:data}.
Techniques for its use are developed in \S \ref{sec:drift} and \S \ref{sec:condit}, leading
in \S \ref{sec:traject} to the study of conditional trajectories in phase space. Finally, a
simple interventional experiment is described in \S \ref{sec:meanprof} to help in the
analysis of the wall regeneration cycle, and conclusions are offered in \S \ref{sec:conc}.

% -----------------------------------------------------------------------------
\section{The Perron--Frobenius operator}\la{sec:PF}

Assume a statistically stationary ergodic system, 
\beq
\bv(t+T)=\bS(T; t) \, \bv(t),
\la{eq:PF0}
\eeq
for which temporal and ensemble averages can be interchanged. The probability density of the
state variable, $\bv$, over the cells of a partition $\{C_j |\, j=1\ldots N\}$ of the phase
space, can be approximated by the fractional distribution, $\bq=\{q_j\}$, of the time spent
by the system within each cell. After a sufficiently long time, or for a sufficiently large
ensemble of experiments, these probabilities tend to an equilibrium distribution that we
denote by $\bq_\infty$. More locally, if we consider the probability distributions at two
different times, $\bq(t)$ and $\bq(t+T)$, the two-dimensional Perron--Frobenius operator
(PFO), $\bhP^e$, relates the past to the future \citep{BeckSchl93},
\beq
\bq(t+T)=\bhP^e(T; t) \, \bq(t).
\la{eq:PF1}
\eeq
Because probabilities represent the results of mutually independent tests, $\bhP^e$ is
linear and, for a finite partition, reduces to an $N\times N$ matrix, where $N$ is the
number of cells in the partition, which is potentially much larger than the number of degrees of
freedom of the original dynamical system. We will assume $\bhP^e$ to be independent of $t$.

When applied to a perfectly concentrated initial distribution, $\bq^{(a)}(t)=\{\delta_{aj}\}$, where
$\delta_{aj}$ is Kronecker's delta, the $a$-th column of $\bhP^e$ represents the
probability that a system initially within the $a$-th cell evolves into the different cells
of the partition after the time interval $T$. Note that these concentrated initial
probability distributions can be interpreted as non-interventional experiments, in which a
statistical knowledge of the causal structure of the coarse-grained system can be gained
by observing the system over a sufficiently long time \citep{AngEtal:96}.

The PFO is equivalent to the Bayesian conditional probability matrix \citep{feller2} and,
after a sufficiently long observation, can be estimated from the joint probability
distribution \citep{Ulam1964}
\beq
\nbQ_{ij}(t, t+T) \equiv \nbQ_{ij}(T) = \mbox{prob}_t(\bv(t+T)\in C_i, \bv(t)\in C_j) ,
\la{eq:PF2d}
\eeq
as
\beq
\nbP^e_{ij} = \nbQ_{ij} /\sum_s \nbQ_{sj},
\la{eq:PF2}
\eeq
This operator is normalised to unit column sums, and an input probability $\bq(t)$,
normalised to $\sum q_j=1$, results in a similarly normalised output probability $\bq(t+T)$. The
matrix $\bP^e$ is generally not symmetric, and there is a dual matrix,
\beq
 {\rm P}^c_{ij} = \nbQ_{ji} /\sum_s \nbQ_{is},
\la{eq:PF3}
\eeq
which generates $\bq(t-T)$ given $\bq(t)$, allowing us to estimate the statistical
distribution of the causes of a given effect. Note that, even if
\beq
\bq(t-T)=\bhP^c(T) \, \bq(t)
\la{eq:PF4}
\eeq
looks like the inverse of \r{eq:PF1}, $\bhP^e$ is not the inverse of $\bhP^c$, because the
marginal probabilities $\bq(t)$ and $\bq(t+T)$ have different meanings in \r{eq:PF1} and
in \r{eq:PF4}. In the former, $\bq(t)$ is observed, and $\bq(t+T)$ is the conditional
probability distribution at $t+T$ given that observation, while their meaning in \r{eq:PF4}
is reversed. One of the effects of the coarse-grained partition is to destroy any
reversibility that might have been present in the original dynamical system.

Another consequence of discrete partitions is to suppress the semigroup character of the
dynamical system, by which $\bS(T_1+T_2)= \bS(T_1)\circ\bS(T_2)$. Indeed, even if
the original dynamical system is Markovian in the sense that its future depends only on its
state at the present (i.e., on its `initial conditions'), the discretised system is
generally not Markovian. The cells of almost any partition of a high-dimensional phase space
are projections of infinite cylinders that extend along some neglected system dimensions.
Two trajectories that intersect a cell at a given time may actually intersect its cylinder
at very different places, and the only way to distinguish different trajectories is often to
consider the sequence of cells visited over their entire past. Even this may not be enough,
and very little is known about partitions that preserve Markovianity in high-dimensional
systems \citep[][\S 3.6]{BeckSchl93}. The transfer operator bypasses this limitation by
acting on the transition probabilities, and is again Markovian in the sense that $\bq(t+T)$
formally only depends on $\bq(t)$ \citep[][\S X]{feller2}, but we regain Markovianity at the
expense of losing determinacy, and we will see in \S\ref{sec:min950} that the semigroup
property, $\bP^e(nT) = \bP^e(T)^n$, is very quickly lost for the approximate transfer
operator of turbulent channels.

There are several reasons why $\bP^e$ is not a perfect estimator of the true
operator $\bhP^e$, but the most important one has to do with the existence of an attractor.
Dissipative systems, such as turbulence, typically evolve towards a lower-dimensional
attracting subset of the full phase space, and the observations used in \r{eq:PF2} only
reflect the statistics of this subset. As such, $\bP^e$ is a restriction of $\bhP^e$ to the
system attractor, and contains little or no information about how the system reacts outside
it. It is thus useful in modelling the physics, where the interest is on how the system
evolves in time, but it may need additional information in control applications, where we
may wish to act in ways outside the attractor.

There are two ways in which the PFO can be used to analyse a complex dynamical system. The
first one is to treat it as a matrix whose properties reflect the behaviour of the attractor
as a whole. `Stochastic' matrices like $\bP^c$ or $\bP^e$, with non-negative elements and
unit column sums, have useful properties that have been extensively studied, especially if
care is exercised in dealing with the zero entries that represent cells that are never visited by
the system \citep{lancaster}. Their best known property is that they posses a unit leading
eigenvalue with a real eigenvector with non-negative entries, which can be interpreted as a
probability distribution over the partition. For $\bP^e$, this eigenvector satisfies, $
\bq_1 (t+T)=\bP^e \bq_1 (t)= \bq_1(t)$, and defines a probability density that remains
invariant as the system evolves, and which is identical to the natural invariant density,
$\bq_\infty$, mentioned earlier in this section. The subdominant eigenvalues control the
approach to $\bq_\infty$ of initial distributions different from the natural one, as well as
whether the attractor can be partitioned into approximately disjoint subsets
\citep{Froy:05}.

As already mentioned, these are examples of global properties that apply to the full
attractor. The same is true of other approximation strategies, such as proper orthogonal
decomposition \citep[POD,][]{Ber:Hol:Lum:93} or dynamic-mode decomposition
\citep[DMD,][]{schmid:10}, which use ergodicity to minimise global errors of reduced models.
We are more interested here in local analyses that use the PFO as a tool for computing
state-dependent conditional averages, and which give information about the expected
short-term behaviour of the system in the neighbourhood of a particular cell. Our goal is to
find whether some cells are more likely than others to form the basis for better
predictions, and are thus `more causal'.

% -----------------------------------------------------------------------------
\section{Application to minimal channels}\la{sec:data}
% ===========================================================
\begin{figure}
\vspace*{5mm}%
\centering
%
%\raisebox{0mm}{\includegraphics[height=.26\textwidth,clip]%
%{\figpath uprof.pdf}}%
%\mylab{-.13\textwidth}{.27\textwidth}{(a)}%
%\hspace*{2mm}%
%%
%\raisebox{0mm}{\includegraphics[height=.26\textwidth,clip]%
%{\figpath vprof.pdf}}%
%\mylab{-.13\textwidth}{.27\textwidth}{(b)}%
%\hspace*{2mm}%
%%
%\raisebox{0mm}{\includegraphics[height=.26\textwidth,clip]%
%{\figpath wprof.pdf}}%
%\mylab{-.13\textwidth}{.27\textwidth}{(c)}%
%
\includegraphics[width=1\textwidth,clip]{\figpath 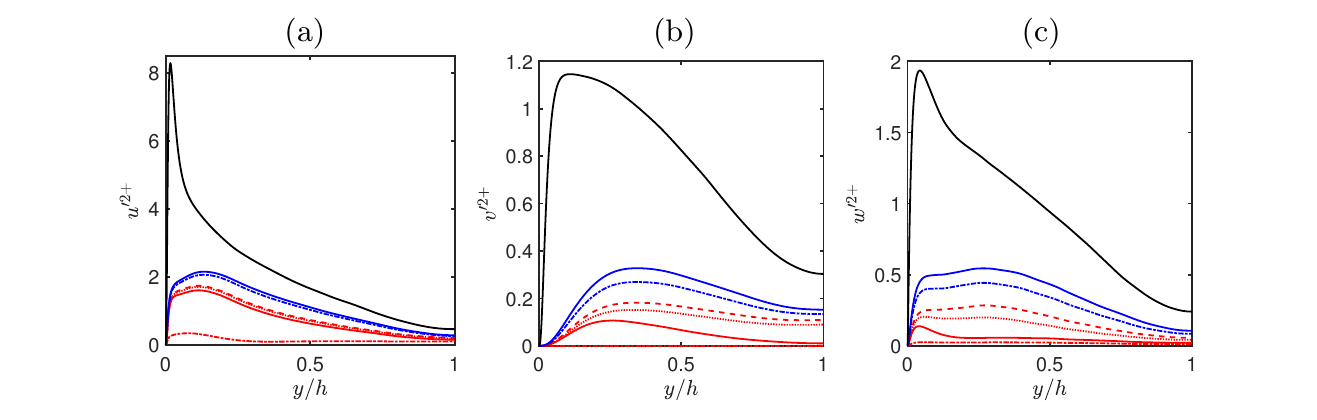}
\caption{%
The coloured lines are profiles of the cumulative variance of the harmonics retained in
this paper. From bottom to top: [00], [01]+[01*], [10], [20], [11]+[11*], [21]+[21*]. The
solid black line is the total variance of the velocity component.
(a) Streamwise velocity. (b) Wall-normal. (c) Spanwise.
}
\label{fig:uprof}
\end{figure}
% ========================================================

Most of our analysis centres on a data set already used in \cite{jimenez:2015}. A
pressure-driven spatially periodic turbulent channel flow is simulated between parallel
plates separated by $2h$. The wall-parallel periods of the computational box are $L_x = \pi
h/2$ and $L_z = \pi h/4$, and the nominal friction Reynolds number is $h^+ = h \utau/\nu=
950$, where $x, y$ and $z$ are the streamwise, wall-normal, and spanwise coordinates,
respectively, and the corresponding velocity components are $u, v$ and $w$. Capital letters
denote $y$-dependent ensemble averages, $\bra\,\ket$, as in $U(y)$, and lower-case ones are
perturbations with respect to this average. Primes are reserved for root-mean-squared
intensities, and the kinetic energy of the fluctuations is defined as $E=u^2+v^2+w^2$. The
`+' superscript denotes `wall' normalisation with the kinematic viscosity $\nu$, and with
the friction velocity $\utau = \sqrt{\nu \p_y U}$. The code is standard fully dealiased
Fourier--Chebychev spectral, as in \cite{kmm}, and the mass flux is kept constant. Time
is usually normalised with the eddy turnoved $h/\utau$, and is then denoted by an asterisk,
$t^*=\utau t/h$. More details can be found in \cite{jim13_lin}.

To improve statistics, the simulation was extended in time to $t^* \approx 650$, and
sampled at a time interval between frames, $\Delta t^* \approx 0.025$. Such simulations
are minimal within a band of wall distances $y/h\approx 0.2 -0.6$
\citep{oscar10_log}, in the sense that a non-negligible fraction of the kinetic energy is
contained in the first few largest wall-parallel Fourier modes. Closer to the wall, the flow
contains a wider range of energy-containing scales, and cannot be considered minimal.
Farther from it, the simulations cannot be directly compared to canonical turbulence,
because some of the largest scales are missing. The range of wall distances mentioned above
approximately includes a single largest structure that bursts irregularly. Since it was
shown by \cite{oscar10_log} that the typical interval between bursts is $t^* \approx 2$--3, the
simulation analysed here contains several hundreds of bursts per wall, and about 100 samples
per burst. Moreover, since the box is too small to allow healthy large scales in the central
part of the channel, the two walls are treated as independent realisations (the
cross-correlation coefficient is less than 0.05 for the variables discussed below).
The total number of data snapshots is thus approximately $5\times 10^4$.

If we define Fourier expansions of the three velocity components along $x$ and $z$ as
\beq
a(x,y,z)=\sum_{m,n} \ta_{mn}(y) \exp [\ii (k_x x +k_z z)] ,
\la{eq:fourdef}
\eeq
where $a$ is the variable to be expanded, $k_x=2\pi m/L_x$, and $k_z=2\pi n/L_z$, the
Fourier coefficients are designated as $[mn]$. As mentioned above, only the largest
structures at a given distance from the wall can be expected to be describable by relatively
few degrees of freedom whose dynamics can be easily studied, and our analysis only retains
the first few modes, $m=0,1, 2$ and $n=-1,0,1$. Appendix \ref{sec:11modes} explains how
modes with $n\ne 0$ are used as combinations of the $\pm n$ pair, resulting in two
equivalent modes displaced spanwise by a quarter of a wavelength. Although spanwise
homogeneity ensures that the interactions of these combinations with the $n=0$ modes are
statistically equivalent, they interact non-trivially among themselves, and both
combinations are retained. They are designated, for example, as [21] and [21*]. Profiles of
the cumulative variance of all the modes used in the paper are given in figure
\ref{fig:uprof}. They show that their overall energy is a comparatively small but
non-trivial fraction of the total, although we will see later that they follow fairly
independent dynamics. In addition, the retained modes account for approximately 65\% of the
tangential Reynolds stress, $-\bra uv\ket$ (not shown). Note that, because of the small
computational box, there is substantial energy in the [00] modes of $u$ and $w$, whose only
fluctuations are temporal. They can approximately be considered as modelling the spatial
variation of the mean velocity profile over wall patches of the order of the size of the
computational box.

This limited subset of data still contains a large number of degrees of freedom, because
each Fourier component is a function of $y$ with $O(100)$ grid points. Even if we will see
later that the wall-normal resolution can be reduced to $O(10)$ points through judicious
filtering, the raw degrees of freedom for each velocity component is $O(100)$ complex
numbers, and we mostly restrict ourselves to analysing the behaviour of a few integrated
`summary variables' that represent global properties of the velocity within the chosen band
of wall distances. In particular, if we are interested in the band $y\in (y_0, y_1)$, we
follow \cite{jim13_lin,jimenez:2015} in using an integrated intensity,
\beq
I^2_{a,mn}=\frac{1}{y_1-y_0}\int_{y_0}^{y_1} |\ta_{mn}^+|^2 \dd y,
\la{eq:ampdef}
\eeq
which stands for the velocity magnitude, and, when $k_x\ne 0$, an average tilting angle 
\beq
\psi_{a,mn}= -\arctan\left({\rm Im}\frac{\int_{y_0}^{y_1} \ta_{mn}^\dag\p_y\ta_{mn}\dd y)}%
            {k_x\int_{y_0}^{y_1} |\ta_{mn}|^2\dd y}\right),
\la{eq:tiltave}
\eeq
where `Im' is the imaginary part, and the dagger stands for complex conjugation. This angle
varies from $-\pi/2$ to $\pi/2$, and describes the wall-normal structure of the phase of the
Fourier mode.

Several other summary variables were considered, either based on physical arguments or in
standard statistical methods (e.g. individual POD modes), but they did not add appreciably
to the argument or to the conclusions. They are not discussed in the rest of the paper,
except for the use of PODs as a filtering device to balance the wall-parallel and
wall-normal resolution of the retained flow fields, as explained in appendix \ref{sec:PODs}.

Because the retained harmonics exclude the smallest scales, they can be trusted closer
to the wall than the full flow, and all our results use an integration band $y^+>40$ and
$y/h \le 0.6$. Somewhat narrower or wider ranges were tested with little effect on the
results.

% ===========================================================
\begin{figure}
\vspace*{5mm}%
\centering
%
%\raisebox{0mm}{\includegraphics[height=.50\textwidth,clip]%
%{\figpath crossall_15X13_nP20.pdf}}%
%
\includegraphics[width=1\textwidth,clip]{\figpath 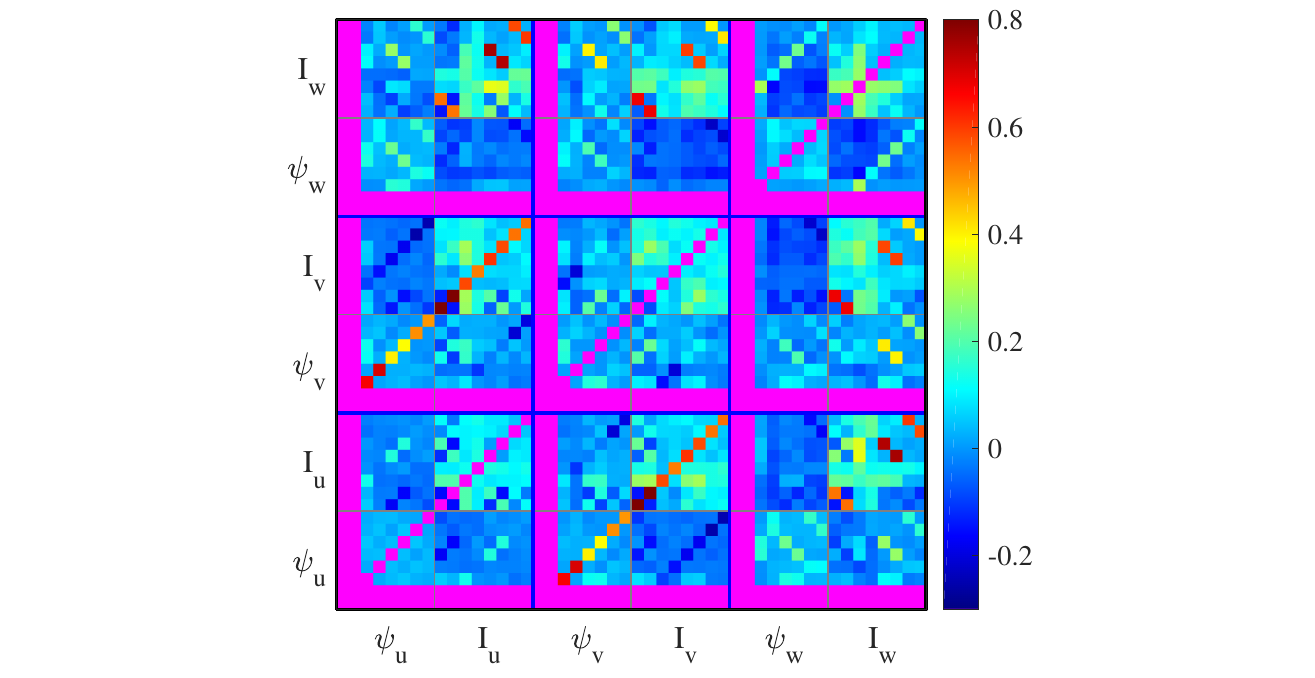}
\caption{%
Correlation coefficient among the different summary quantities. Large squares outlined in
blue correspond to the three velocity components. Smaller squares outlined in grey are
summary variables, and the smallest cells within each grey square are Fourier modes, in the
order [01], [01*], [10], [20], [11], [11*], [21], [21*], from bottom to top and from left to
right. The main diagonal has been blocked for clarity, as well as the inclinations for modes with
$k_x=0$, which are undefined.
}
\label{fig:crosscorr}
\end{figure}
% ========================================================

Not all the summary variables defined in this way are mutually fully independent.
Figure \ref{fig:crosscorr} presents their correlation coefficient,
\beq
C_{ab}= \frac{ \bra (a-\bra a\ket)  (b-\bra b\ket) \ket}
      {\bra (a-\bra a\ket)^2 \ket^{1/2}\,  \bra (b-\bra b\ket)^2 \ket^{1/2}}.
\la{eq:corrce0}
\eeq
Several things stand out. The $u$ and $v$ components form reasonably well-correlated pairs,
particularly among similar summary variables and Fourier modes, but most quantities
involving $w$ are not well correlated with $u$ and $v$, or among themselves. The correlation
between the intensities of $u$ and $v$ are significant because, even if they involve
integrated quantities rather than the variables themselves, they reflect the generation of
the tangential Reynolds stress, $-uv$. The higher modes, [11] and [21], tend to be better
correlated among different variables than the lower ones, [10] and [20]. In particular,
three of the highest correlations in figure \ref{fig:crosscorr} are $C(I_{v01},
I_{w01*})\approx 0.70$, $C(I_{u11}, I_{w11*})\approx 0.74$, and $C(I_{v11}, I_{w11*})\approx
0.59$, notwithstanding the generally poor correlation between the summaries of $w$ and those
of other velocity components. Interestingly, these correlations come in [$m$1, $m$1*] pairs,
representing flow structures offset from each other by a quarter of a spanwise wavelength.
They correspond to the inclined `rollers' that have often be described in wall-bounded
flows.

Somewhat surprisingly, angles and intensities are generally uncorrelated, including the $(I_{v10},
\psi_{v10})$ pair that was shown by \cite{jim13_lin,jimenez:2015} and \cite{encinar:20} to
be particularly useful, because its joint probability distribution is traversed by the flow
in a physically interpretable way. This shows that correlations and coherence are different
concepts. As a simple example, the temporal evolutions of $\sin(t)$ and $\cos(t)$ are
orthogonal and uncorrelated, but they form a coherent pair in the sense that they transverse a
one-dimensional circular sub-manifold of their phase space.

It could be tempting to use as summary variables the eigenvectors of the dominant
eigenvalues of the matrix in figure \ref{fig:crosscorr}. These combinations of variables
optimally explain the variance of the data \citep{Ber:Hol:Lum:93}, but they turn out to be
especially bad at describing the dynamics. This can best be understood by looking at the
joint probability density of $(I_{v10},\psi_{v10})$ in figure \ref{fig:PFangamp}(a). It is
clear that $I_{v10}$ does not explain $\psi_{v10}$, nor vice versa, which is precisely why
the pair can be used to define two-dimensional causal combinations.

We explore in the rest of the paper whether other interesting pairs of variables can be
found.

% .........................................................................
\subsection{The transfer operator of the minimal channel}\la{sec:min950}
 
% ===========================================================
\begin{figure}
\vspace*{5mm}%
%
%\centerline{%
%%
%\raisebox{0mm}{\includegraphics[height=.27\textwidth,clip]%
%{\figpath invdvp4_va4_t3_15X13_nP20.pdf}}%
%\mylab{-.18\textwidth}{.28\textwidth}{(a)}%
%\hspace*{3mm}%
%%
%\raisebox{0mm}{\includegraphics[height=.27\textwidth,clip]%
%{\figpath PFvp4_va4_t3_15X13_nP0.pdf}}%
%\mylab{-.175\textwidth}{.280\textwidth}{(b)}%
%%
%\hspace*{3mm}%
%%
%\raisebox{0mm}{\includegraphics[height=.27\textwidth,clip]%
%{\figpath Markovvp4_va4_15X13_nP0.pdf}}%
%\mylab{-.14\textwidth}{.280\textwidth}{(c)}%
%%\hspace*{6mm}%
%}%
\includegraphics[width=1\textwidth,clip]{\figpath 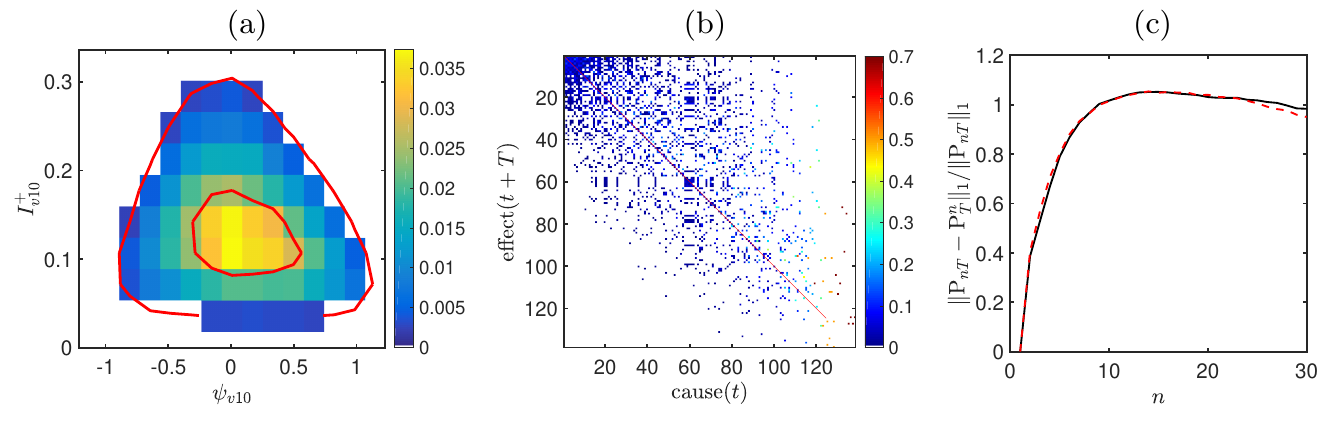}
\caption{%
(a) Two-dimensional joint probability distribution, $\bq_\infty$, for the integrated inclination
and amplitude of the [10] mode of the wall-normal velocity component 
(i.e., $k_x=2\pi/L_x , k_z=0$), averaged over $y^+>40$ and $y/h<0.6$. The inclination is
partitioned in 15 equal bins, and the amplitude in 13 bins. The red contours contain 30\%
and 95\% of the probability mass, and only cells within the outer contour are plotted.
(b) PF operator, $\bP^e$, obtained by collating the variables in (a) into a single vector,
offset by $T^*=0.078$, and rearranged in order of decreasing column sum of the joint
probability $\bQ$ in \r{eq:PF2d}. Rows and columns with zero sum have been eliminated.
(c) $L_1$-norm Markov test for the PF matrices. \solid, $\bP^e$; \dashed, $\bP^c$.
}
\label{fig:PFangamp}
\end{figure}
% ========================================================

It should be clear from the previous discussion that the main problem in constructing the
PFO for a given system is the choice of the underlying partition, and of the variables in
which it is expressed. Figure \ref{fig:PFangamp} presents results for the minimal channel
just described, using as variables the inclination angle of the wall-normal velocity,
$\psi_{v10}$, and its root-mean-squared amplitude, $I_{v10}$, whose joint probability
distribution is shown in figure \ref{fig:PFangamp}(a). Most of the distribution is contained
within the inner probability contour, but the outer fringe is interesting because
\cite{jimenez:2015} showed that its upper edge can be modelled as a linearised burst in
which the mean shear amplifies the velocity perturbations by tilting them forward
\citep{orr07a,jim13_lin}, although the process by which the cycle is closed is less
well understood.

The construction of the PFO starts by organising the $15\times 13$ partition of the
parameter space of figure \ref{fig:PFangamp}(a) into a single vector of length 195, and
constructing the two-time joint distribution, $\bQ(t,\,t+T)$, from all the snapshots in the
data sequence. The interval used in figure \ref{fig:PFangamp}, $T^*=0.078$, is chosen
from the experience in \cite{jimenez:2015} and \cite{encinar:20}, and is the time taken
by the system to traverse an increment  $\Delta\psi \approx 0.3$ along the upper edge of figure
\ref{fig:PFangamp}(a). It is also the time over which \cite{jimenez:2015} shows that the
flow can be linearly predicted in that region of the phase plane.

The columns with non-zero elements in $\bQ$ are normalised using \r{eq:PF2}, and those with
only zeros are discarded. The resulting estimated $\bP^e$ is shown in figure
\ref{fig:PFangamp}(b), where each column is normalised to unity, and sorted in order of
decreasing row sum for graphical purposes. The restriction of the full $\bhP^e$ to the
on-attractor estimate $\bP^e$ is done when discarding zeros at this step.

Note that, although the discretised variables are treated as parts of a single long vector,
the probabilities in $\bQ$ correspond to the simultaneous occurrence of two independent
variables, allowing us to study effects depending on two `causes' \citep{Pearl:09}. The
procedure could be extended to three- or higher-dimensional projections of phase space, and
to the effects of the increasingly unlikely coincidence of three or more independent causes,
but it quickly runs into the limitation of the number of available data. The $5\times 10^4$
snapshots used here have to populate the $138^2$ matrix $\bP^e$ in figure
\ref{fig:PFangamp}(b), giving an average of 2.5 phase points per matrix element. In
practice, they range from $O(100)$ data for the better populated matrix elements to zero for
elements outside the attractor. The limitation is not as strict for the smoother
distribution in figure \ref{fig:PFangamp}(a), which is essentially an eigenvector of the PFO
(see \S\ref{sec:PF}). Its dimension is just 138 cells, and each cell represents
$O(300$--1000) data points. Tests with partitions of the order of $10\times 10$ cells did
not qualitatively change the results described below, but attempts to use much finer
partitions ran into problems at the interesting edge of the distribution.

As a consequence, we restrict ourselves to two-dimensional projections of the phase space,
and explore in \S \ref{sec:drift} which pairs of summary variables give more interesting results.
This is supplemented in \S\ref{sec:condit} by conditionally averaging other variables over
these projection planes, recovering part of the three-dimensional dynamics.

In addition, to ensure that the noise in our results is a consequence of the discrete
partition rather than of an insufficient number of data, the distribution in figure
\ref{fig:PFangamp}(a), and similar later ones, are only drawn within the probability
isocontour containing 95\% of the total probability mass. Each cell along this contour contains
$O(100)$ data snapshots. The global averages in \S \ref{sec:drift} are also computed within this
high-probability region, and the analysis was repeated with half the number of data, with
similar conclusions.

% ===========================================================
\begin{figure}
\vspace*{5mm}%
\centering
%
%\centerline{%
%%
%\raisebox{0mm}{\includegraphics[height=.33\textwidth,clip]%
%{\figpath caueffvp4_va4_t3_15X13_nP20_5_7.pdf}}%
%\mylab{-.26\textwidth}{.29\textwidth}{(a)}%
%\hspace*{11mm}%
%%
%\raisebox{0mm}{\includegraphics[height=.33\textwidth,clip]%
%{\figpath caueffvp4_va4_t3_15X13_nP20_8_4.pdf}}%
%\mylab{-.26\textwidth}{.29\textwidth}{(b)}%
%\hspace*{10mm}%
%}%
%%
%\vspace{2mm}%
%%
%\centerline{%
%%
%\raisebox{0mm}{\includegraphics[height=.33\textwidth,clip]%
%{\figpath Efflowvp4_va4_t3_15X13_nP20.pdf}}%
%\mylab{-.32\textwidth}{.29\textwidth}{(c)}%
%\hspace*{2mm}%
%%
%\raisebox{0mm}{\includegraphics[height=.33\textwidth,clip]%
%{\figpath Cauflowvp4_va4_t3_15X13_nP20.pdf}}%
%\mylab{-.32\textwidth}{.29\textwidth}{(d)}%
%}%
%\vspace{2mm}
%%
%\centerline{%
%%
%\raisebox{0mm}{\includegraphics[height=.33\textwidth,clip]%
%{\figpath Efflowvp4_va4_t3_15X13_nP20_rand.pdf}}%
%\mylab{-.31\textwidth}{.29\textwidth}{(e)}%
%\hspace*{2mm}%
%%
%\raisebox{0mm}{\includegraphics[height=.33\textwidth,clip]%
%{\figpath Cauflowvp4_va4_t3_15X13_nP20_rand.pdf}}%
%\mylab{-.31\textwidth}{.29\textwidth}{(f)}%
%}%
%
\includegraphics[width=1\textwidth,clip]{\figpath 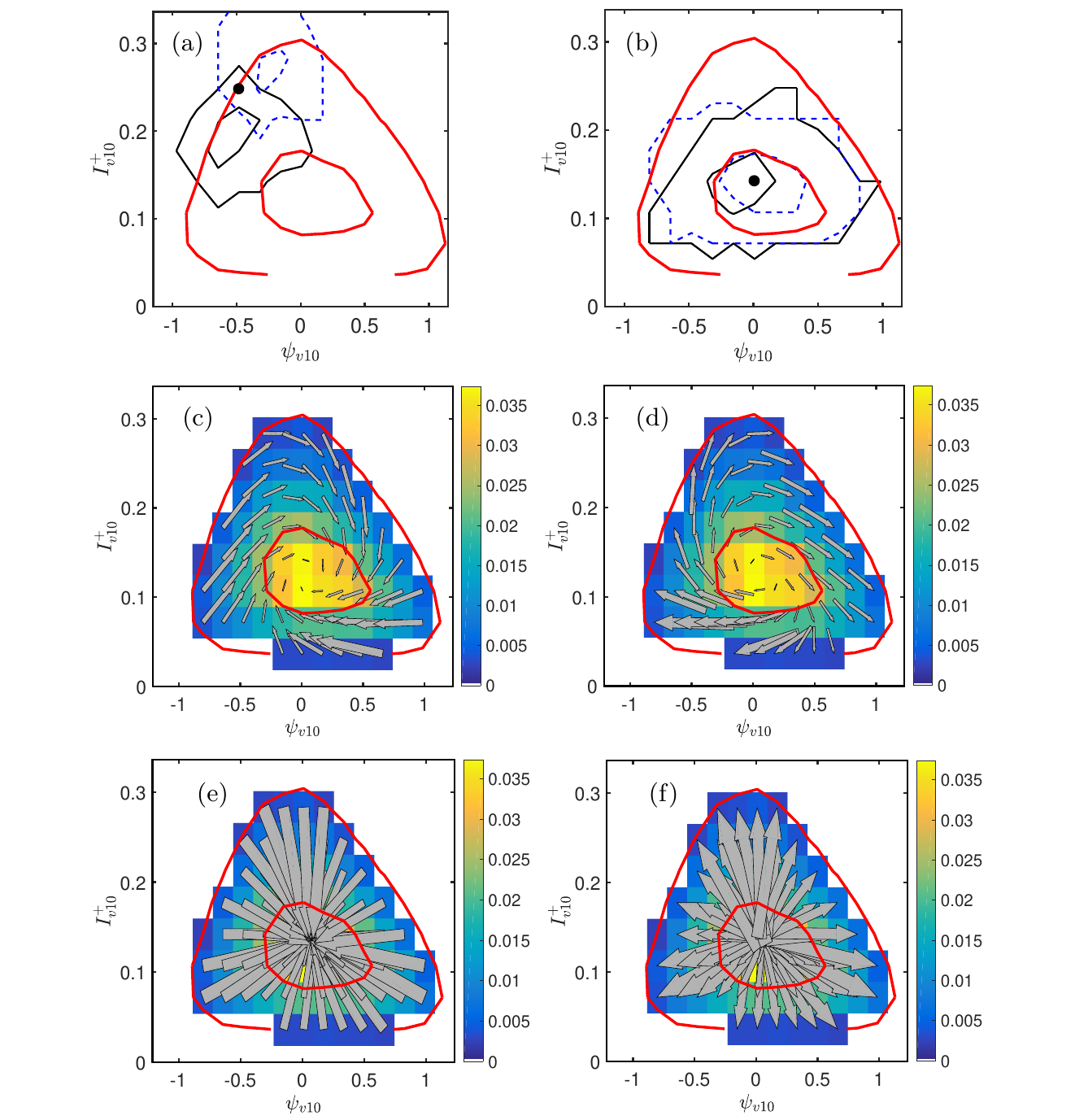}

\caption{%
(a) For the variables in figure \ref{fig:PFangamp}(a), and an interval $T^*=0.078$, the
solid black contours are the probability distribution of possible $T$-precursors to an
observation of the cell marked with a solid circle, and the dashed blue contours are the
distribution of possible effects after $T$. Contours contain 30\% and 95\% of the
probability mass.
(b) As in (a), for an observation in the core of the invariant density distribution. 
(c) Mean system displacements in the parameter plane. The coloured background is the
invariant density, and the arrows join the cell taken as cause with the mean system location after
time $T$.
(d) As in (c), but the arrows join the mean location of the systems that will pass through
the cell taken as reference after time $T$. The red contours contain 30\% and 95\% of the
invariant density.
(e,f) As in (c,d), but using randomised time stamps for the data.
}
\label{fig:caueff}
\end{figure}
% ========================================================

Figure \ref{fig:PFangamp}(c) tests the non-Markovian behaviour of $\bP^c$ and $\bP^e$
discussed in \S \ref{sec:PF}. It shows the relative Frobenius norm of the difference
between $\bP(nT)$ and $\bP(T)^n$ for $T^*= 0.025$. The relative difference between
uncorrelated stochastic matrices depends on the ratio between the standard deviation and the
mean of individual matrix columns, but is approximately unity for cases such as those in figure
\ref{fig:PFangamp}. It is clear that the two matrices being tested become essentially
uncorrelated after a few time steps, and from now on we use $\bP(nT)$ as our basic operator. 
 
Figure \ref{fig:caueff} shows how the PFO can be used to extract the probability
distributions of the causes and effects of a given observation. Figure \ref{fig:caueff}(a)
assumes that we know that the system is within the cell marked with a
solid circle at $t=0$. The conditional probability distribution at $t=T$ is given by the
corresponding column of the transfer operator, $\bP^e$, and is displayed in the figure in
dashed blue contours. Conversely, the conditional probability distribution of 
causes at $t=-T$ is the corresponding column of the backwards operator $\bP^c$. It is
displayed in solid black lines, and the difference among the two distributions illustrates the
temporal evolution of the system in the clockwise direction of the figure, as in
\cite{jimenez:2015}.

The segregation into forward and backward distributions does not hold for all cells. Figure
\ref{fig:caueff}(b) applies the procedure to a cell in the high-probability core of the
invariant density distribution. Its forward and backward distributions are marked as in
figure \ref{fig:caueff}(a), but they overlap each other and are difficult to tell apart.

Figure \ref{fig:caueff}(c) is a representation of this mean displacement for all the cells
in the distribution. The arrows join the centre of each reference cell to the mean position
of its effects after a given time interval. Figure \ref{fig:caueff}(d) does the same for the
causes, and both figures show a mean clockwise displacement of the system along the upper
edge of the distribution \citep[see figure 3b in][for comparison]{jimenez:2015}. In addition
to this circular displacement, the arrows spiral towards the centre of the distribution in
figure \ref{fig:caueff}(c), and outwards in figure \ref{fig:caueff}(d). This tendency
increases for longer time intervals, and is due to the non-Markovian component of the
probability evolution.

Any random displacement from the periphery tends to move towards the most probable
locations in the central part of the distribution, and random displacements into the periphery
are most likely to come from the core. This is best seen in figures \ref{fig:caueff}(e,f),
which are computed in the same way as figures \ref{fig:caueff}(c,d) after randomising the
time stamps of the flow snapshots. In fact, since the effects and causes are in this case
randomly chosen states of the system, their expected average coincides with the overall mean
of the invariant distribution. These randomised figures are independent of the time interval.

% ----------------------------------------------------------------------------------------------
\subsection{Quality indicators}\la{sec:drift}

We have seen in the previous section that the main problem in constructing the PFO is collecting
enough data to populate the two-dimensional histogram, $\bQ$, making it unpractical to
consider distributions over more than two independent variables. We have also mentioned that
our strategy is to test all possible variable pairs in the hope of identifying couples whose
statistical behaviour is optimal, but the 36 variables used in figure \ref{fig:crosscorr}
can be paired into 630 possible combinations, and automating the search requires indicators
that are simpler to implement than the visual inspection of the two-dimensional plots in
figure \ref{fig:caueff}. Four such indicators are discussed in this section.

% ===========================================================
\begin{figure}
%\vspace*{5mm}%
%\centerline{%
%%
%\raisebox{0mm}{\includegraphics[height=.33\textwidth,clip]%
%{\figpath Separvp4_va4_t3_15X13_nP20.pdf}}%
%\mylab{-.31\textwidth}{.29\textwidth}{(a)}%
%\hspace*{2mm}%
%%
%\raisebox{0mm}{\includegraphics[height=.33\textwidth,clip]%
%{\figpath corcevp4_va4_t1_15X13_nP20.pdf}}% 
%\mylab{-.31\textwidth}{.29\textwidth}{(b)}%
%}%
%%
%\vspace*{3mm}%
%%
%\centerline{%
%\raisebox{0mm}{\includegraphics[height=.33\textwidth,clip]%
%{\figpath Hellvp4_va4_t3_15X13_nP20.pdf}}%
%\mylab{-.31\textwidth}{.29\textwidth}{(c)}%
%\hspace*{2mm}%
%%
%\raisebox{0mm}{\includegraphics[height=.328\textwidth,clip]%
%{\figpath KLincvp4_va4_t3_15X13_nP20.pdf}}%
%\mylab{-.31\textwidth}{.29\textwidth}{(d)}%
%%
%}%
%
\includegraphics[width=1\textwidth,clip]{\figpath 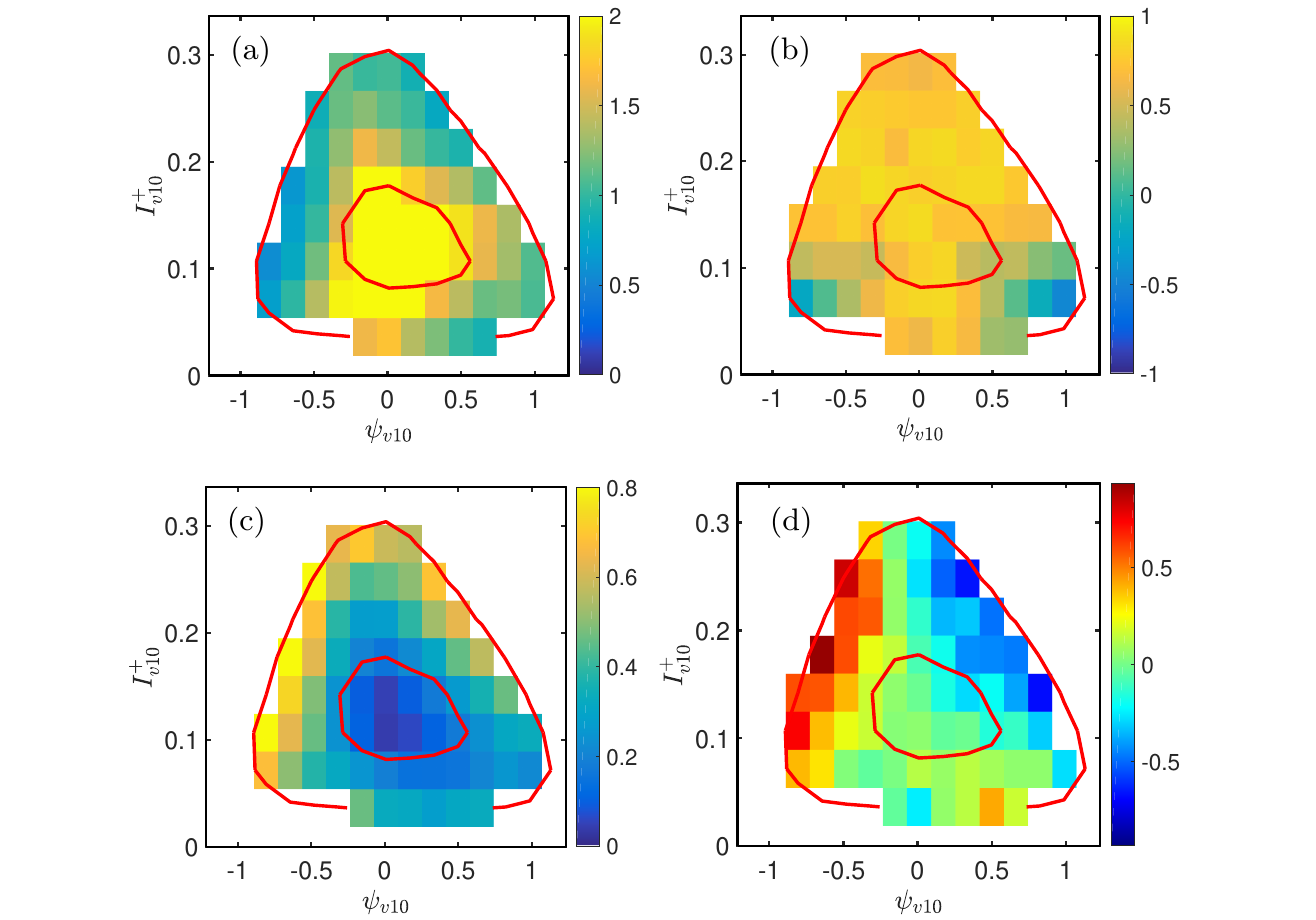}
\caption{%
As in figure \ref{fig:PFangamp}.
(a) Ratio between the averaged displacement of the effects and their standard deviation. Each
variable is normalised with its overall standard deviation to compensate for the different
units.
(b) Determinacy index \r{eq:corrce} between the average displacement of causes and effects.
Drawn for $T^*=0.025$.
(c) Hellinger segregation index \r{eq:Hell} between the forward and backwards conditional
distributions, as function of the observation cell.
(d) Kullback--Leibler information gain \r{eq:KL2} from the distributions of the effects
and  the causes, measured in bits. Warm colours represent creation of information, and cooler
ones represent information loss.
All panels refer to the [10] mode of the wall-normal velocity, and use only cells within
the 95\% probability contour of $\bq_\infty$. In all of them, except (b), $T^*=0.078$.
}
\label{fig:entropies}
\end{figure}
% ========================================================

The statistical uncertainty of the displacement vectors is addressed in figure
\ref{fig:entropies}(a), which displays the ratio between the standard deviation of the
conditionally averaged displacement of the system over a given time and its mean. To
compensate for the different magnitudes of the two variables in the figure, which generally
have incompatible units, each of them is normalised with its global standard deviation
before computing the conditional statistics. The result is a measure of the error bars
associated with each of the arrows in figure \ref{fig:caueff}(c).

Having a small relative standard deviation does not guarantee that a quantity is physically
relevant. Inspection of figure \ref{fig:caueff}(c-f) reveals that deterministic and random
evolutions behave differently with respect to the asymmetry between causes and effects. The
displacement vectors of the causes and effects rotate in the same direction in figure
\ref{fig:caueff}(c,d), because both represent the deterministic evolution of the system. But
the randomised vectors in figures \ref{fig:caueff}(e,f) point in opposite directions,
because they move from the conditioning cell towards the densest part of the distribution,
independently of the direction of time. As a consequence, we can define a `determinacy'
index for an observation cell $\bv_0$ as the normalised inner product
\beq
C_{ce}(\bv_0, T)= \frac{ (\bv^e-\bv_0)\cdot (\bv_0-\bv^c)}
      {\|\bv^e-\bv_0\| \|\bv_0-\bv^c\|},
\la{eq:corrce}
\eeq
where $\bv^e-\bv_0$ and $\bv_0-\bv^c$ are, respectively, the conditionally averaged
displacement of effects and causes over the time interval $\pm T$. As in figure
\ref{fig:entropies}(a), variables are normalised with their standard deviation before
computing \r{eq:corrce}. This index is an indication of how deterministic is the
evolution of the system in the neighbourhood of $\bv_0$, and of how much information is
gained by the observation of the variable pair. When the system is completely deterministic
in the subspace being considered, $C_{ce}\approx 1$, and when it is essentially random,
$C_{ce}\approx -1$. Figure \ref{fig:entropies}(b) displays $C_{ce}$ for the data in figure
\ref{fig:caueff}, and shows that the evolution of this particular Fourier mode in this parameter
plane is deterministic almost everywhere.

Along the upper edge of the distribution, this agrees with the physically based conclusions
of \cite{jimenez:2015}, but not along its lower edge, where both \cite{jimenez:2015} and
\cite{encinar:20} conclude that the average displacement is opposite to the predictions of
the model that explains the upper edge, and that the uncertainty of the displacements is too
large to trust their mean. The high uncertainty in this region is clear in figure
\ref{fig:entropies}(a), but figure \ref{fig:entropies}(b) suggests that this part of the
distribution is also deterministic. Part of the reason is the longer time interval used in
figure \ref{fig:entropies}(a) compared to \ref{fig:entropies}(b). The apparent randomness of
the evolution increases for longer intervals, as the non-Markovian behaviour takes over. The
determinacy index is almost unity in figure \ref{fig:entropies}(b), where the displacements
are of the order of one distribution cell, but decreases to $C_{ce}\approx 0.8$ when the
figure is drawn for the more physically relevant time interval used in figure
\ref{fig:entropies}(a), and to $C_{ce}\approx 0.5$ for the even longer interval used in
\cite{jimenez:2015}. \cite{encinar:20}, who use a different method from the one above, and a
different set of data, compute a figure of merit equivalent to the relative dispersion in
figure \ref{fig:entropies}(a). Normalising their time offset with the average distance, $\overline y$, from
the wall of their filtered fields \citep{oscar10_log,jimenez:2015}, it varies
between $\utau T/\overline y = 0.048$ and 0.19. The resulting standard deviations
are negligible for the shortest of those intervals, but large enough to reverse some of
the displacements for the largest one. When these values are applied to the present case,
assuming $\overline y\approx 0.3$ for our integration band, the time interval in figure
\ref{fig:entropies}(b) is $\utau T/\overline y = 0.087$, and that in figure
\ref{fig:entropies}(a) is $\utau T/\overline y = 0.26$, explaining the apparent discrepancy
between figures \ref{fig:entropies}(a) and \ref{fig:entropies}(b).
 
Figure \ref{fig:entropies}(c) quantifies the temporal segregation between the conditional
probability distributions of causes and effects in figure \ref{fig:caueff}(a,b). The
distance between two normalised probability distributions $\bq^{(1)}$ and $\bq^{(2)}$ can be
characterised by the Hellinger norm \citep{nikulin01}, defined as
\beq
H^2(\bq^{(1)},\bq^{(2)}) = \tfrac{1}{2} \sum_j \left(\sqrt{q^{(1)}_j}-\sqrt{q^{(2)}_j}\right)^2 ,
\la{eq:Hell}
\eeq
which vanishes for $\bq^{(1)}=\bq^{(2)}$, and reaches its maximum, $H=1$, for disjoint
distributions. In the case of figure \ref{fig:caueff}(a,b) and \ref{fig:entropies}(c), the
distance between the conditional distributions of causes and effects varies from $H\approx
0.9$ at the edge of the density distribution, where they are clearly different, to
$H\approx 0.1$ at the centre, where past and future are almost indistinguishable.

The information provided by the indices \r{eq:corrce} and \r{eq:Hell} is related but not
identical. While a high value of \r{eq:Hell} implies that causes and effects are
different, a high value of \r{eq:corrce} also shows that the directions of the mean drift
associated to each of them are similar, and that the flow of probability can be described as a smooth
vector field.

When figures \ref{fig:entropies}(a-c), are considered together they suggest that the
top-right and top-left edges of the probability distribution are populated by systems which
evolve in fairly deterministic manner, while the lower edge of the distribution, and especially its central core,
are more random.

Finally, figure \ref{fig:entropies}(d) addresses the question of whether this 
evolution has any effect in the probability distribution of the variables used in this
section. In essence, whether the effects conditioned to a given cell are more or less
organised than its causes. The Kullback-Leibler (KL) information of a distribution
$\bq^{(1)}$, relative to a reference distribution $\bq^{(2)}$, is defined as
\beq
K(\bq^{(1)},\bq^{(2)}) = \sum_j q^{(1)}_j \log_2 (q^{(1)}_j/ q^{(2)}_j),
\la{eq:KL1}
\eeq
which is measured in bits, is always non-negative, and only vanishes when
$\bq^{(1)}=\bq^{(2)}$. Intuitively, it describes how much more organised is $\bq^{(1)}$
compared to $\bq^{(2)}$. Note that \r{eq:KL1} is only finite if the support of $\bq^{(1)}$
is contained within the support of $\bq^{(2)}$, so that $K$ can also be understood as a
measure of how much information is gained by restricting $\bq^{(2)}$ to one of its subsets.
Here, we will always use as reference the invariant distribution $\bq_\infty$, so that $K$ is
guaranteed to exist both for the distribution $\bq^c$ of the conditional causes and for the
distribution $\bq^e$ of the effects. This choice also implies that a distribution with $K=0$
is statistically indistinguishable from $\bq_\infty$, and represents an unconstrained set of
phase points. The assumption that the system is restricted to a single cell at $t=0$ almost
guarantees that information is lost when this concentrated distribution is allowed to spread
in the past or in the future, but the information contained in the two distributions cannot
be compared directly, because they do not generically share a common support. Figure
\ref{fig:entropies}(d) displays the difference,
\beq
K^{ce} =  K(\bq^e,\bq_\infty)-K(\bq^c,\bq_\infty),
\la{eq:KL2}
\eeq
between the information of conditional effects and of conditional causes with respect to the
reference. It is positive along the left (growth) edge of the distribution, and negative
along the right (decay) edge, suggesting that coherence is first created and later destroyed
as the system drifts clockwise. Because the system is stationary, the two effects cancel,
and the mean generation of information vanishes.

% ===========================================================
\begin{figure}
%\vspace*{6mm}%
\centering
%
%\raisebox{0mm}{\includegraphics[height=.41\textwidth,clip]%
%{\figpath deterall_15X13_nP20.pdf}}%
%\mylab{-.275\textwidth}{.425\textwidth}{(a)}%
%%
%\hspace{3mm}%
%%
%\raisebox{0mm}{\includegraphics[height=.41\textwidth,clip]%
%{\figpath Hellall_15X13_nP20.pdf}}%
%\mylab{-.275\textwidth}{.425\textwidth}{(b)}%
%
\includegraphics[width=1\textwidth,clip]{\figpath 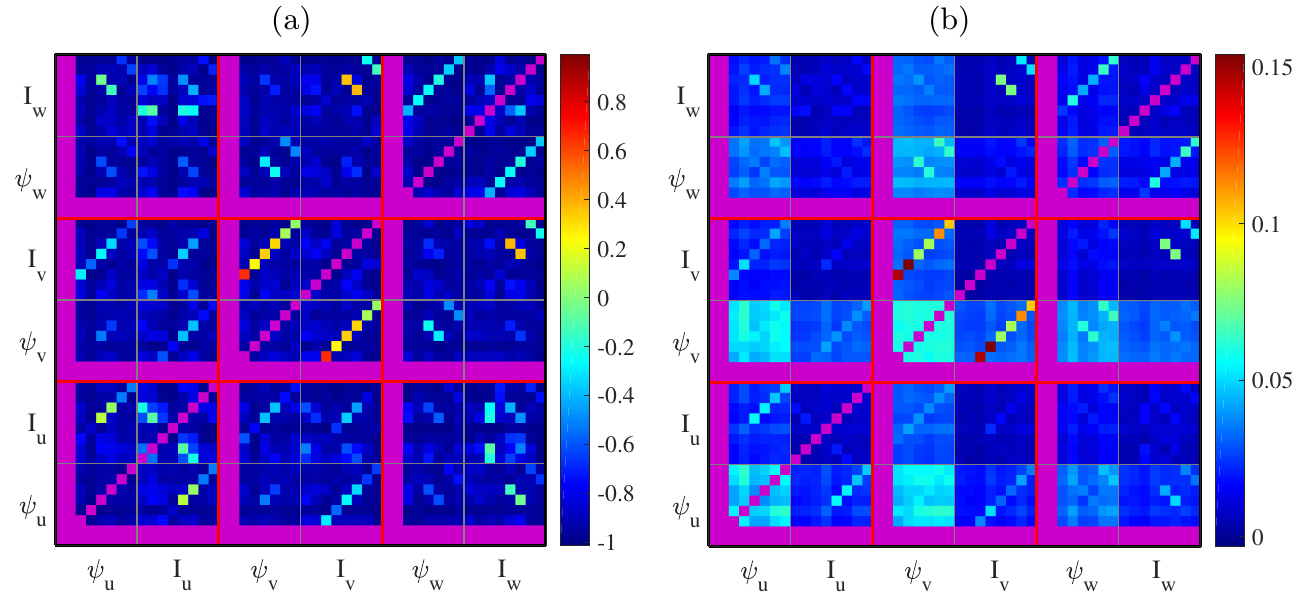}
\caption{%
As in figure \ref{fig:crosscorr}. 
(a) Determinacy index \r{eq:corrce}, averaged over the invariant distribution for different
combinations of modal inclination and intensity. 
(b) Segregation index \r{eq:Hell}. The main diagonal and the inclinations of the $k_x=0$
modes are blocked to magenta in both cases.
}
\label{fig:allcorrel}
\end{figure}
% ========================================================

However the real power of the indicators is in the large-scale screening of less obvious
variable pairs. Figure \ref{fig:allcorrel} displays the value of the determinacy
\r{eq:corrce} and segregation \r{eq:Hell} indices for all the combinations of modal
inclination and intensity of the three velocity components, averaged over the corresponding
distributions. Not surprisingly, the most deterministic combinations are those involving the
inclination and intensity of the same Fourier mode, but it is interesting that $v_{10}$,
which was the mode selected on physical grounds in \cite{jimenez:2015}, is also the most
deterministic by a substantial margin. Slightly less deterministic is $v_{20}$, which is
just a harmonic of $v_{10}$ for which the same theory applies. The meandering modes,
$v_{11},\,v_{21},\,u_{11},\,u_{21}$, are less organised but not fully
incoherent. The spanwise velocity is not well described by an inclination and an intensity,
but an interesting pair is $(I_{v11}, I_{w11*})$, or vice versa, which was already
identified as an inclined roller from the correlations in figure \ref{fig:crosscorr}. Its
appearance in the two indicators in figure \ref{fig:allcorrel} shows that this roller has
its own causal dynamics, and that the same is true for the second harmonic, $(I_{v21},
I_{w21*})$. However, the streamwise-uniform roller $(I_{v01}, I_{w01*})$ does not appear in
figure \ref{fig:allcorrel}, even if it is one of strongest couplings in the correlations in
figure \ref{fig:crosscorr}, and one of the largest contributors to the fluctuation energies
in figure \ref{fig:uprof}. Such two-dimensional structures do not interact with the shear
and, even if they may grow to be strong, have little own dynamics.

% ===========================================================
\begin{figure}
%\vspace*{1mm}%
%%
%\centerline{%
%%
%\raisebox{0mm}{\includegraphics[height=.31\textwidth,clip]%
%{\figpath Efflowva6_wa7_t3_15X13_nP20.pdf}}%
%\mylab{-.105\textwidth}{.27\textwidth}{(a)}%
%\hspace*{0.5mm}%
%%
%\raisebox{0mm}{\includegraphics[height=.315\textwidth,clip]%
%{\figpath corcenoyva6_wa7_t1_15X13_nP20.pdf}}%
%\mylab{-.105\textwidth}{.27\textwidth}{(b)}%
%%
%\hspace*{0.5mm}%
%%
%\raisebox{0mm}{\includegraphics[height=.31\textwidth,clip]%
%{\figpath KLincnoyva6_wa7_t3_15X13_nP20.pdf}}%
%\mylab{-.105\textwidth}{.27\textwidth}{(c)}%
%%
%}%
\includegraphics[width=1\textwidth,clip]{\figpath 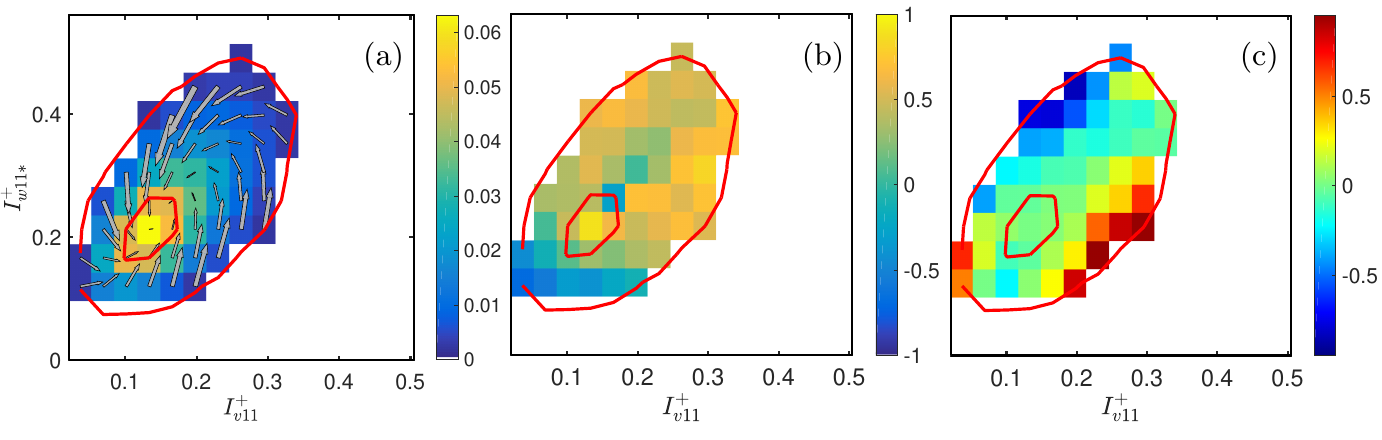}
\caption{%
As in figures \ref{fig:caueff} and \ref{fig:entropies}, for the 
roller variables, $(I_{v11}, I_{w11*})$.
(a) Quiver plot of the effects, as in figure \ref{fig:caueff}(c).
(b) Determinacy index \r{eq:corrce}, as in figure \ref{fig:entropies}(b). 
(c) Kullback--Leibler information gain \r{eq:KL2}, as  in figure \ref{fig:entropies}(d). 
}
\label{fig:morentropies}
\end{figure}
% ========================================================

Figure \ref{fig:morentropies} displays the drift diagram and two quality indicators for the
 $(I_{v11}, I_{w11*})$ roller. The two variables are relatively well
correlated, probably by continuity, and the distribution rotates counter-clockwise. This
implies that the roller is first created as an ejection (or sweep) of the wall-normal
velocity, and later spreads spanwise. Figure \ref{fig:morentropies}(c) shows that, as in
figure \ref{fig:entropies}(d), coherence is created when the structure strengthens, and
destroyed when it weakens.

% -----------------------------------------------------
\subsection{Conditional averages}\la{sec:condit}

\begin{figure}
\includegraphics[width=1\textwidth,clip]{\figpath 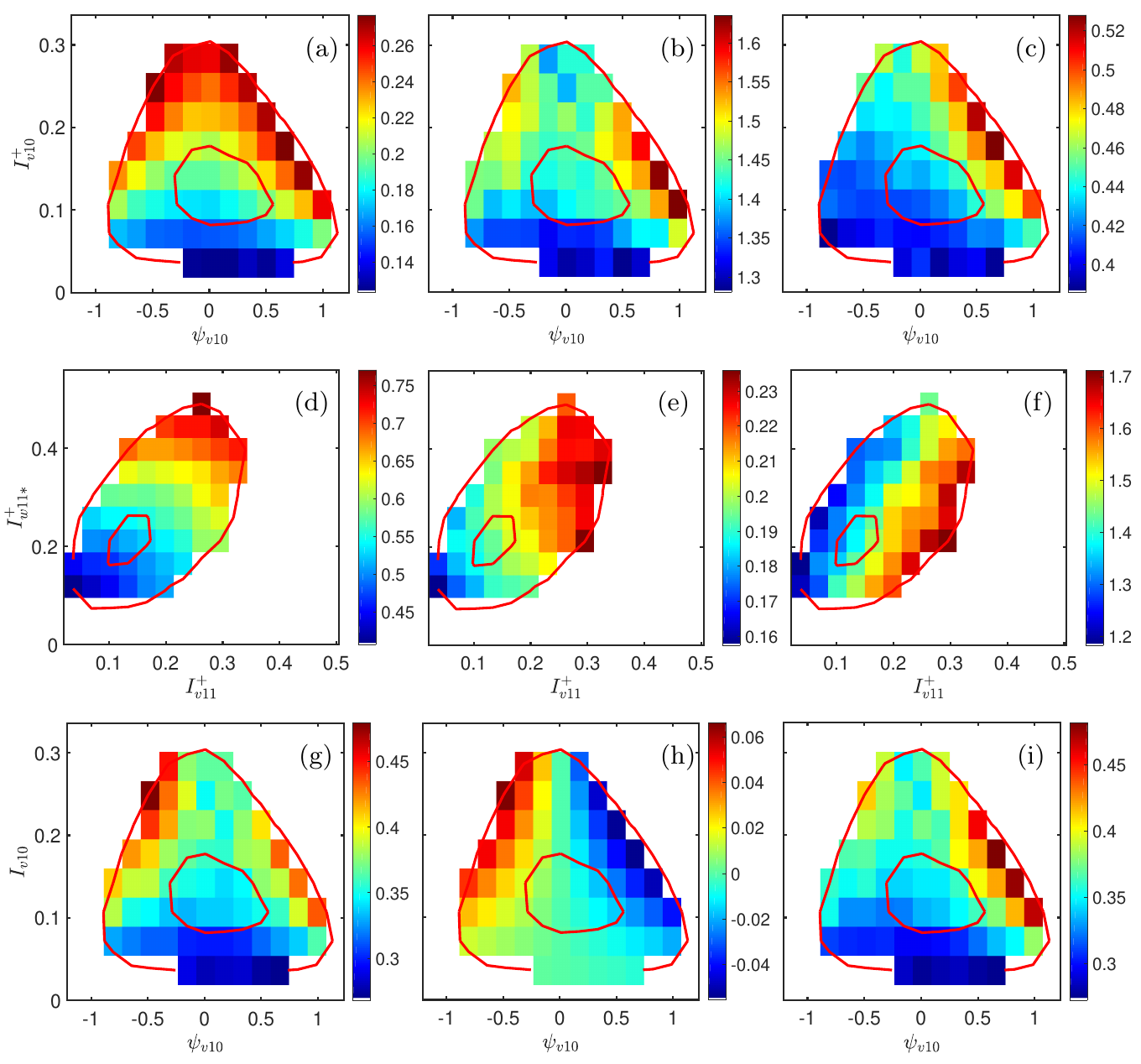}
\caption{%
(a-c) Average intensity of different velocity modes, conditioned to the variable pair $(\psi_{v10},I_{v10})$.
(a) $u_{10}$. (b) All modes of $u$ with $k_z\ne 0$. (c) $\sqrt{v^2_{11}+w^2_{11}}$.
(d-f) Conditioned to $(I_{v11}, I_{w11*})$.
(d) $u_{11}$. (e) $u_{10}$. (f) All modes of $u$ with $k_z\ne 0$.
(g-i) Conditional tangential Reynolds stress, $\theta$. 
(g) All the retained harmonics. (h) $\theta_{10}$. (i) All harmonics with $k_z\ne 0$.
}
\label{fig:condit}
\end{figure}

The analysis in the previous sections reveals which variable pairs evolve in a
deterministic way, but says little about the associated flow fields. A first
step in that direction is figure \ref{fig:condit}, which displays the averages of other
flow variables conditioned to the basic  $(\psi_{v10},I_{v10})$ pair. Most
conditional averages provide relatively little information. They are either distributed
almost uniformly over the invariant probability distribution, or track the evolution of the
base variables, saying little more than that strong structures are strong in most variables.
A few are more interesting.

Figure \ref{fig:condit}(a) shows that the streamwise velocity becomes more non-uniform in
the streamwise direction $(u_{10})$ as the wall-normal velocity bursts. It follows from the
choice of conditioning variables that this also holds for the wall-normal velocity,
but is not true of the spanwise component (not shown). The coupling of $u_{10}$ and $v_{10}$
is required by continuity, and the two variables are part of the two-dimensional burst
described in \cite{orr07a} and \cite{jim13_lin}. However, self-sustaining turbulence
requires three-dimensionality, and this is provided by the $k_z\ne 0$ modes in figure
\ref{fig:condit}(b). This figure displays the modes of $u$ that form a possibly non-uniform
streak. They grow along the decay leg of the burst, and are therefore probably 
consequences, rather than precursors of the burst. It can be shown that most of the growth in
figure \ref{fig:condit}(b) is due to the [01] mode, which measures the intensity of a
streamwise-uniform streak, and is also responsible for most of the asymmetry of the figure
along the $\psi$ axis. The higher modes, [11] and [21], only grow weakly, and do so along
the ascending leg of the burst. On the other hand, the cross-flow $(v,w)$ roller grows
along the descending branch, as shown in figure \ref{fig:condit}(c) for the [11] harmonic.
This is also true for the [01] streamwise roller, which has similar intensity, but much less
for the [21] mode, which is weaker and less coherent. In all these cases, the two signs of
$k_z$ are grouped in the figure.

% ===========================================================
% ========================================================

When the intensity of the $x$-dependent modes is substituted by a non-uniformity index,
\beq
\Omega_{a,mn} =I_{a,mn}/I_{a,0n},
\la{eq:waviness}
\eeq
figure \ref{fig:condit} changes somewhat. The growth of the roller in figure
\ref{fig:condit}(c) is transferred to the bottom of the distribution. Its non-uniformity
begins with $w_{11}$ in the descending leg of the burst, and moves to $u_{11}$
and $v_{11}$ along its left-going bottom edge. The wavy quasi-streamwise roller is very weak
in that region, but also very non-uniform, and it is tempting to conclude that this waviness
is part of what eventually triggers a subsequent burst.

Figure \ref{fig:condit}(d-f) shows conditional averages in the plane of the $(I_{v11},
I_{w11*})$ meandering roller. Figure \ref{fig:condit}(d) is equivalent to figure
\ref{fig:condit}(a), and confirms that the roller is a three-dimensional structure in which
the streamwise velocity increases as $v_{11}$ and $w_{11*}$ do. The effect is much stronger
for $u_{11}$ than for $u_{11*}$, showing that the active streamwise velocity is collocated
with the wall-normal component, $v_{11}$, rather than with $w_{11*}$. The
three-dimensionality extends to other harmonics, and figure \ref{fig:condit}(e) shows that
one effect is to enhance the streamwise non-uniformity of the spanwise-uniform component
$u_{10}$. This figure is dual to figure \ref{fig:condit}(a), and suggests that the
intensification of the roller takes place along the upper edge of the $(\psi_{v10},I_{v10})$
burst, although where exactly cannot be decided from this representation. For example, the
conditional $I_{v10}$, which would be a direct indicator of the position along the
$(\psi_{v10},I_{v10})$ burst, does not produce a clear signal in the $(I_{v11}, I_{w11*})$
plane. Figure \ref{fig:condit}(f) shows the conditional $k_z\ne 0$ component of $u$, which
is the same quantity in figure \ref{fig:condit}(b), and gives more information. Figure
\ref{fig:condit}(b) shows that the streak of $u$ grows along the descending leg of the
clockwise evolution of $(\psi_{v10},I_{v10})$, while figure \ref{fig:condit}(f) shows that
it grows along the ascending leg of the counter-clockwise evolution of $(I_{v11},
I_{w11*})$. Both legs therefore presumably correspond to the same stage of the flow, in
agreement with the location in figure \ref{fig:condit}(c) of the high roller intensity.
 
Also interesting are figures \ref{fig:condit}(g-i), which show the conditional tangential
Reynolds stress, $\theta_{mn}=-{\rm Re} (\tu_{mn} \tv^\dag_{mn})$, integrated as in
\r{eq:ampdef}. Although the quantity in the equations of motion is $\p_y\theta$, rather than
$\theta$ itself, a positive stress tends to make the mean velocity profile more turbulent,
steeper near the wall, and negative ones tend to lower the wall shear. Figure
\ref{fig:condit}(g) displays the conditional Reynolds stress due to all the retained
harmonics, which we saw in \S\ref{sec:data} to account for approximately two thirds of the
total flow stress. It is positive everywhere, and stronger in the upper edge of the
distribution, where other flow features are also strong. It is also asymmetric in the
$(\psi_{v10},I_{v10})$ plane, stronger during the growth of the burst than along its decay.
The reason is shown in figure \ref{fig:condit}(h), which displays the stress due to the
\cite{orr07a} harmonic, $\theta_{10}$. It is almost antisymmetric in $\psi_{v10}$, positive
during the growth of the burst, and negative during its decay. This is what makes the
burst transient, since its decay undoes the effect of the growth. The [10] mode is the only
one that generates counter-gradient stresses. Figure \ref{fig:condit}(i) displays the
tangential stress from all the other modes. They are positive everywhere, and the net effect
of the burst, although transient in itself, is to generate a three-dimensional structure
along its decay leg.

Although not shown in the figure, it is interesting that, when $\theta_{10}$ is conditioned
to the $(I_{v11}, I_{w11*})$ plane, it is also negative in the uppermost tip of the roller
distribution, confirming our previous conclusion that strong rollers correspond to the burst decay.

In a loose sense, both the down-going right-hand edge of the probability distribution in
figure \ref{fig:condit}(a-c) and the up-going right-hand edge of figure
\ref{fig:condit}(d-f), portray the evolution of a non-uniform $u$-streak into a $(v,w)$
roller. In the top row of figure \ref{fig:condit} we see the decay of the streak, and in the
middle one we see the growth of the roller. The comparison of the conditioned variables in
both sets of figures suggests that the coherent right-hand edge of the evolution of the
roller corresponds to the decay of the burst, so that the approach to the lower-right vertex
of the triangle in figure \ref{fig:condit}(a-c) corresponds to the top of the distribution
in figure \ref{fig:condit}(d-f). In this interpretation, the low-intensity evolution of the burst along
the bottom edge of the distribution in figures \ref{fig:condit}(a-c) should correspond in
part to the decay of the roller along the left edge of the distribution in figures
\ref{fig:condit}(d-f). This phase of the burst will be examined in more detail in the next section.

% ---------------------------------------------------------------------------------------
\section{Conditional trajectories}\la{sec:traject}

% ===========================================================
\begin{figure}
%\vspace*{5mm}%
%\centerline{%
%%
%\raisebox{0mm}{\includegraphics[height=.403\textwidth,clip]%
%{\figpath Orbhisvp4_va4_15X13_nP20.pdf}}%
%\mylab{-.05\textwidth}{.34\textwidth}{(a)}%
%\hspace*{7mm}%
%%
%\raisebox{0mm}{\includegraphics[height=.40\textwidth,clip]%
%{\figpath trayecvp4_va4_t0_t26_ip7_ia11_ig6923_15X13_nP20.pdf}}%
%\mylab{-.15\textwidth}{.34\textwidth}{(b)}%
%}%
%%
%\vspace*{5mm}%
%\centerline{%
%%
%\raisebox{0mm}{\includegraphics[height=.40\textwidth,clip]%
%{\figpath trayecva6_wa7_t0_t26_ip7_ia11_ig6923_15X13_nP0.pdf}}%
%\mylab{-.14\textwidth}{.34\textwidth}{(c)}%
%%
%\hspace*{1mm}%
%\raisebox{0mm}{\includegraphics[height=.40\textwidth,clip]%
%{\figpath trayecvp4_va4_t0_t26_ip7_ia11_all_15X13_nP0.pdf}}%
%\mylab{-.14\textwidth}{.34\textwidth}{(d)}%
%}%
%
\includegraphics[width=1\textwidth,clip]{\figpath 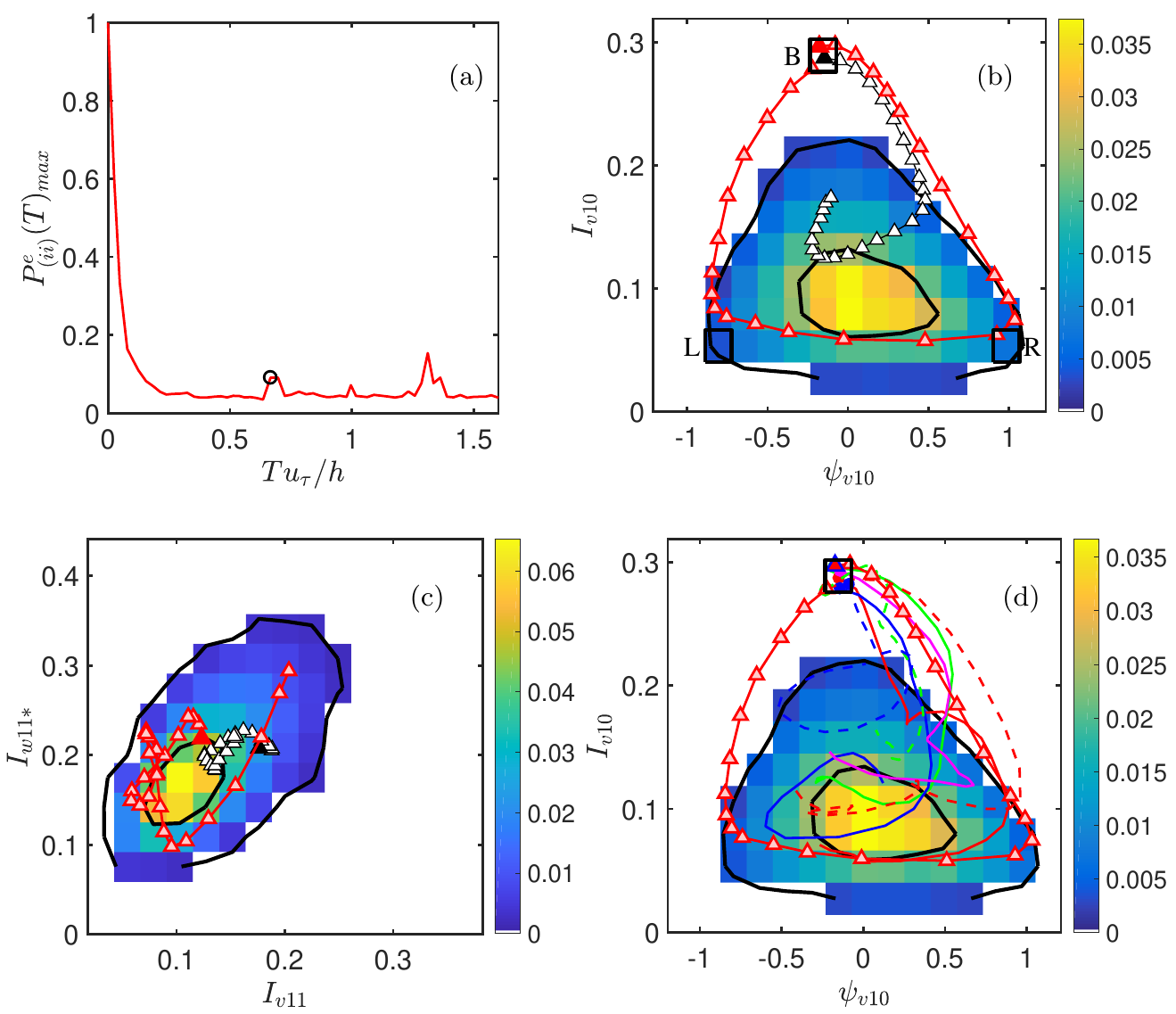}
\caption{%
(a) Recurrence test for trajectories in the $(\psi_{v10},I_{v10})$ plane. See text for explanation.
(b) Black symbols are the mean trajectory passing through the most probable recurrent cell,
for the period marked by the dark circle in (a). The red symbols are the only recurrent orbit crossing
that cell, which is thus responsible for the local maximum in (a).
(c) As in (b), in the  $(I_{v11}, I_{w11*})$ plane of the meandering roller.
(d) As in (b) but all the non-recurrent trajectories are plotted as simple lines, each one
starting with a solid circle. 
}
\label{fig:orbit}
\end{figure}
% ========================================================

It may be useful at this point to recall that the maps in figures such as \ref{fig:condit}
are not the PFO, but its leading eigenvector rearranged as a two-dimensional matrix for
human convenience. The operator itself is the larger matrix in figure \ref{fig:PFangamp}(b),
or a stack of such matrices at different time intervals. The maps in figure \ref{fig:condit}
are probability distributions of the state of the flow in phase space, and the
delta-function distribution used to generate figure \ref{fig:caueff}(a,b)
represents a measurement that collapses that probability to a single combination of
variables that describes a flow configuration. For the rest of the paper we will study these
collapsed probabilities, and the statistical properties of the trajectories connecting two or
more such states. The PFO then becomes mostly a guide to which cells can be expected to
provide more interesting information, and to how the trajectories connecting them should be
chosen.

For example, consider the statistical characterisation of trajectories that
satisfy conditions at two different times, such as in the classical recurrence
test in which approximate periodic behaviour is identified by monitoring when a trajectory
approaches itself after a given delay \citep{KawEtal12}. The statistical equivalent is
whether the conditional probability distribution of the effects of a given cell includes the
conditioning cell after some delay. In practice, this reduces to identifying among the
diagonal of the matrix $\bP^e(T)$ those cells that result in maximum probability of
recurrence for each time delay. An example is figure \ref{fig:orbit}(a), which plots the
maximum of the diagonal of the PFO for the $(\psi_{v10},I_{v10})$ plane. It is unity at $T=
0$, when trajectories are still at their initial position, and quickly decays to about 0.05,
which is the probability that a random trajectory intersects some cell in the core of the
invariant distribution $\bq_\infty$. However, the curve peaks again at $ T^* = 0.66$,
and, interestingly, at twice that delay, $T^* = 1.35$. Moreover, since the PFO
contains information about all the cells in the distribution, it allows us to recover which
conditioning cell is responsible for the probability maximum, and therefore which cell has
the highest probability of recurring. This is marked as `B' (for burst) in figure
\ref{fig:orbit}(b), and turns out to be an extreme high-amplitude event beyond the 95\%
threshold used up to now as the practical edge of our distribution. In the trajectories
discussed in this section, closed symbols mark the position at $t=0$, and the open triangles
along some trajectories are equispaced by $t^*=0.025$.

Two other cells are labelled in figure \ref{fig:orbit}(b), marking the right- (`R') and
left-hand (`L') corners of the triangular distribution. We will maintain this nomenclature
for the rest of the section, with some adjustments in the location of the cells.
Trajectories spanning the up-going (R $\to$ B), down-going (B $\to$ L), and bottom (L $\to$
R) legs of the periphery of the triangle will be denoted as growth, decay and recovery
trajectories, respectively. The growth and decay legs form the burst \citep{jim13_lin}. The
upper half of these two legs is deterministic, and can be predicted linearly
\citep{jimenez:2015}, but linearised bursts do not recur. There is no obvious theory for the
bottom recovery leg, which is required if bursting is to explain self-sustaining turbulence.
Most of this section is dedicated to analysing the recovery process.

Out of our $5\times 10^4$ snapshots, only eight trajectories cross the extreme B
in figure \ref{fig:orbit}. Most of them do not recur, and the line of open black triangles
in figure \ref{fig:orbit}(b) traces the average conditional trajectory during the recurrence
period. As is true for most trajectories, it approaches the high-probability core of the
distribution. However, individual trajectories can be tested, and the line of red triangles
in figure \ref{fig:orbit}(b) shows the trajectory responsible for the peak in figure
\ref{fig:orbit}(a). Centring ourselves on this orbit, its growth, decay and recovery legs
last approximately $T^*=0.25,\, 0.25$ and 0.16, respectively, for a total recurrence time
$T^* \approx 0.66$, as in figure \ref{fig:orbit}(a). The total length of its two bursting
legs, $T^* \approx 0.5$, also agrees with the width of the bursting correlations for this
flow in \cite{jimenez:2015}.

Although the recurrent trajectory very approximately closes on itself in the plane of figure
\ref{fig:orbit}(b), its recurrence is weaker when more variables are included. For example,
figure \ref{fig:orbit}(c) plots the same trajectories in the plane of the $(I_{v11},
I_{w11*})$ roller. As before, the recurring trajectory is displayed as red triangles. It
loiters for a while near the high-probability core of the $(I_{v11}, I_{w11*})$
distribution, and joins the counter-clockwise coherent circulation
during the growth period of the $(\psi_{v10},I_{v10})$ burst. The black triangles of the
mean trajectory are again very different from the recurrent one, and never join the coherent
circulation.

However, the divergence between the recurrent trajectory and the other trajectories that
cross B is mostly a long-time phenomenon. This can be seen in figure \ref{fig:orbit}(d),
which is equivalent to figure \ref{fig:orbit}(b) but plots all the individual trajectories
going through B. The recurrent orbit is still plotted with open red triangles, while other
trajectories are plotted as lines of different colours without symbols, in no particular
order. All the trajectories in the figure start from the same cell and behave similarly for
a while. It is only after they have decayed to approximately half their initial $I_{v10}$
that they deviate towards the high-probability core. A similar plot in the $(I_{v11},
I_{w11*})$ plane, not shown, shows the same trends, although slightly more complicated
because trajectories do not start in the same cell any more. The recurrent orbit is especial
in that it starts with a relatively weak and decaying roller, which only
strengthens towards the end of the orbit. This is probably the reason why its
$(\psi_{v10},I_{v10})$ projection is able to proceed undisturbed for a relatively long time.

% ===========================================================
\begin{figure}
%\vspace*{5mm}%
%\centerline{%
%%
%\raisebox{0mm}{\includegraphics[height=.40\textwidth,clip]%
%{\figpath traileffvp4_va4_t1_6_15X13_nP20_14_2.pdf}}%
%\mylab{-.08\textwidth}{.34\textwidth}{(a)}%
%\hspace*{1mm}%
%%
%\raisebox{0mm}{\includegraphics[height=.40\textwidth,clip]%
%{\figpath trailcauvp4_va4_t1_6_15X13_nP20_3_2.pdf}}%
%\mylab{-.08\textwidth}{.34\textwidth}{(b)}%
%}%
%
\includegraphics[width=1\textwidth,clip]{\figpath 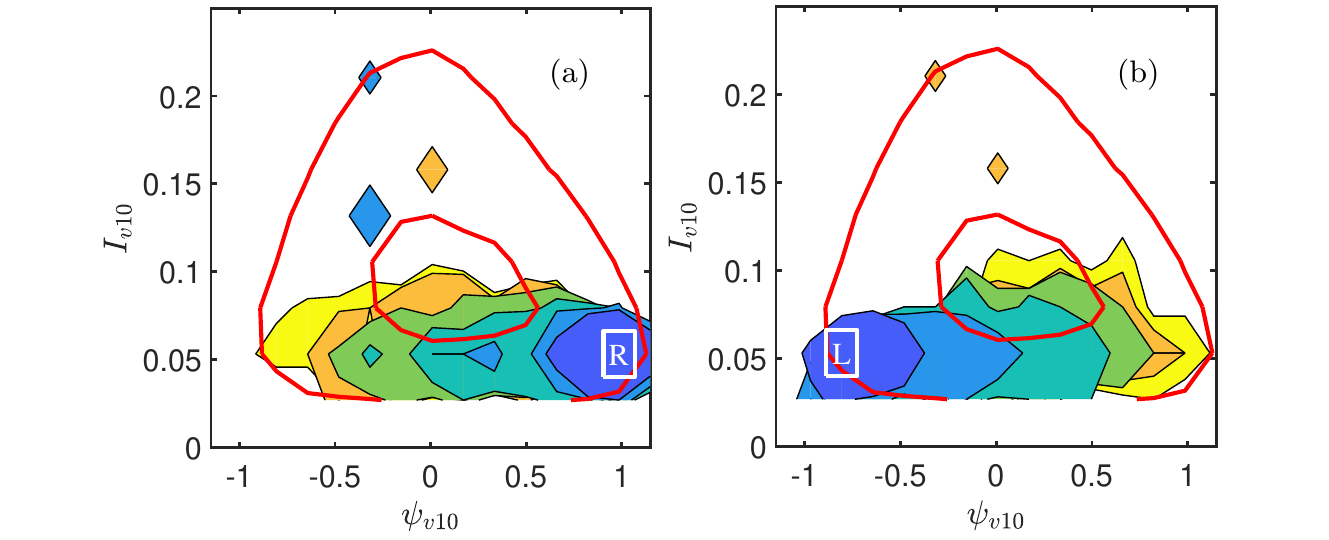}
\caption{%
Probability distributions of the causes and effects for several delay intervals, as in
figure \ref{fig:caueff}(a,b). (a) Effects conditioned to the cell marked as R in the lower
right-hand corner of the invariant density distribution. From blue to yellow, $T^* =
0.025 (0.025) 0.15$. (b) As in (a), for the causes leading to the cell marked as L in the
lower left-hand corner.
}
\label{fig:lowedge}
\end{figure}
% ========================================================

That the approximately recurrent orbit includes an infrequent extreme event suggests that it
may not be very relevant to the flow statistics, or even to its evolution. Its
interest resides in that it includes a recovery leg that traverses the lower edge of the
parameter space in the `wrong' direction, showing that coherent recovery
connections exist, and suggesting how to study them. The question is whether other
trajectories exist that cross the lower edge of the distribution in the same direction, even
if they do not follow the recurrent orbit around the full $(\psi_{v10},I_{v10})$ plane.

\begin{figure}
\includegraphics[width=1\textwidth,clip]{\figpath 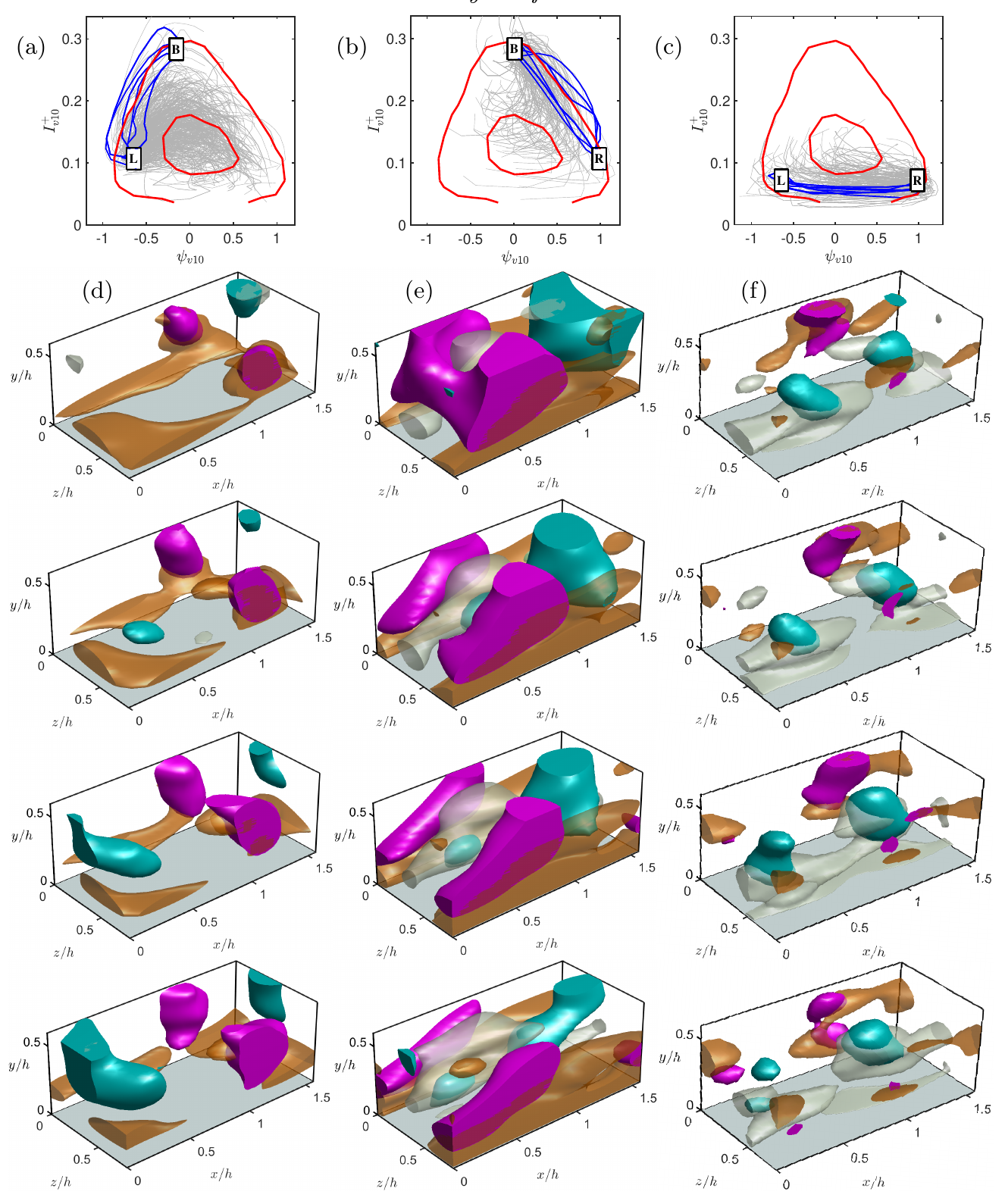}
\caption{%
(a-c) Grey lines are the phase trajectories that cross a causal cell at $t=0$. Blue
ones are those that also cross the effect cell within a given range of time intervals.
(a) Growth leg of the burst, from L $\to$ B in $T^* = 0.2$ -- 0.3.
(b) Decay from B $\to$ R in $T^* = 0.2$ -- 0.3.
(c) Recovery leg from R $\to$ L in $T^* = 0.15$ -- 0.175.
(d-f ) Averaged evolution of the flow along the blue trajectories in (a-c), respectively, 
reconstructed from the retained harmonics.
Flow is from left to right, and time from top to bottom.
Translucent orange isosurface, $u^+=0.7$; translucent grey, $u^+=-0.7$; 
cyan, $v^+=0.4$; purple, $v^+=-0.4$. 
(d-e) $t^*=0.075 (0.05) 0.225$.
(f)  $t^*=0 (0.05) 0.15$, and  $v^+=\pm 0.3$.
}
\label{fig:legs}
\end{figure}

Figures \ref{fig:lowedge} and \ref{fig:legs} address this question. Figures
\ref{fig:lowedge}(a,b) are equivalent to the cause and effect distributions in figure
\ref{fig:caueff}(a), but applied to the R an L corner cells for several values of the delay
interval. Even if we saw in figure \ref{fig:entropies}(a) that the lower edge of the
distribution is a region of large scatter for the system drift, figure \ref{fig:lowedge}(a)
shows that the flow propagates steadily from R towards L as the delay increases, and that it
only spills slowly into the high probability core of the distribution. The same is true for
the causes of cell L in figure \ref{fig:lowedge}(b). When similar plots are drawn for a cell
in the central part of the lower edge of the distribution, the drift is slower and the
scatter higher (not shown), suggesting that the propagation in figure \ref{fig:lowedge}(a,b)
can be interpreted as a direct connection between R and L that bypasses some of the
intermediate states of the probability distribution.

% ===========================================================
% ========================================================

% ========================================================
\begin{figure}
%\vspace*{1mm}%
%\centerline{%
%%
%\raisebox{0mm}{\includegraphics[width=.32\textwidth,clip]%
%{\figpath uavv_vp4_va4_t8_12_15X13_nP20_4_3_7_8.pdf}}%
%\mylab{-.05\textwidth}{.20\textwidth}{(a)}%
%\hspace*{2mm}%
%%
%\raisebox{0mm}{\includegraphics[width=.32\textwidth,clip]%
%{\figpath uavv_vp4_va4_t8_12_15X13_nP20_8_8_14_3.pdf}}%
%\mylab{-.05\textwidth}{.20\textwidth}{(b)}%
%\hspace*{2mm}%
%%
%\raisebox{0mm}{\includegraphics[width=.32\textwidth,clip]%
%{\figpath uavv_vp4_va4_t6_7_15X13_nP20_14_2_4_2.pdf}}%
%\mylab{-.26\textwidth}{.20\textwidth}{(c)}%
%}%
%%
%\vspace*{1mm}%
%\centerline{%
%%
%\hspace*{0.01\textwidth}\raisebox{0mm}{\includegraphics[width=.31\textwidth,clip]%
%{\figpath Ev_vp4_va4_t8_12_15X13_nP20_4_3_7_8.pdf}}%
%\mylab{-.05\textwidth}{.20\textwidth}{(d)}%
%\hspace*{2mm}%
%%
%\hspace*{0.01\textwidth}\raisebox{0mm}{\includegraphics[width=.31\textwidth,clip]%
%{\figpath Ev_vp4_va4_t8_12_15X13_nP20_8_8_14_3.pdf}}%
%\mylab{-.05\textwidth}{.20\textwidth}{(e)}%
%\hspace*{2mm}%
%%
%\hspace*{0.01\textwidth}\raisebox{0mm}{\includegraphics[width=.31\textwidth,clip]%
%{\figpath Ev_vp4_va4_t6_7_15X13_nP20_14_2_4_2.pdf}}%
%\mylab{-.26\textwidth}{.20\textwidth}{(f)}%
%}%
%
\includegraphics[width=1\textwidth,clip]{\figpath 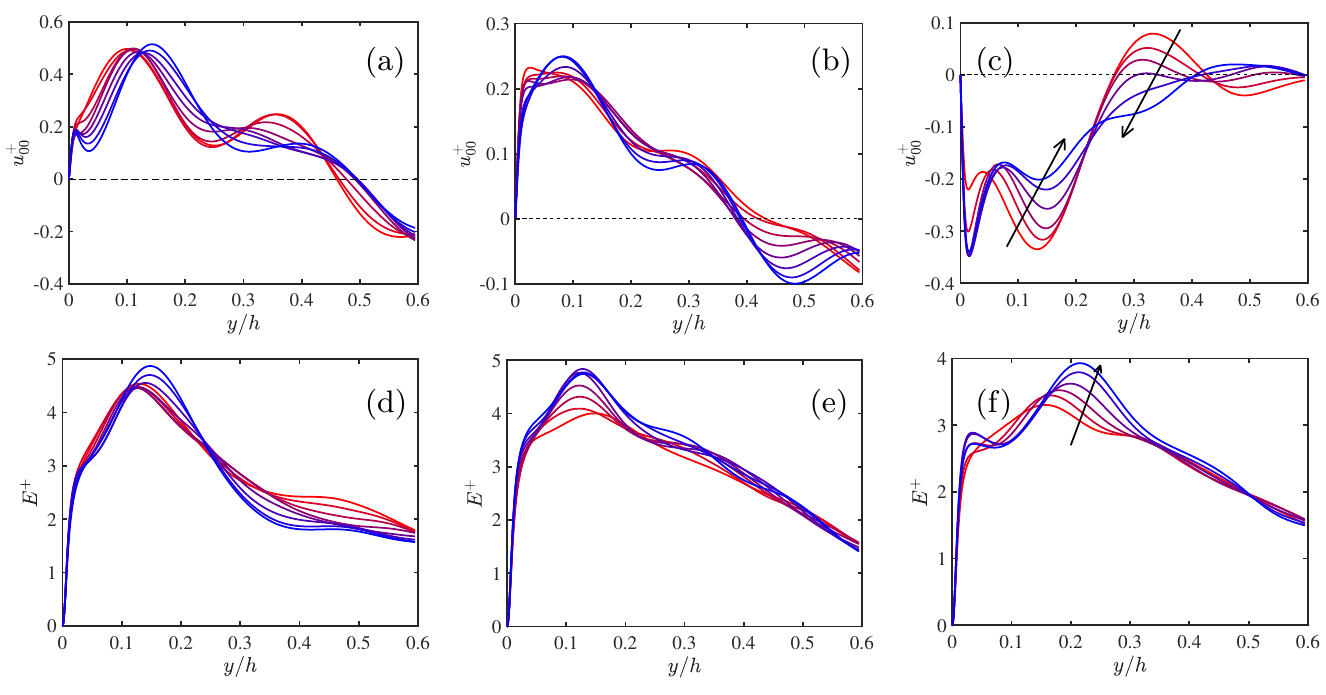}
\caption{%
Box-averaged velocity profiles, for the trajectories marked in blue in figure \ref{fig:legs}. 
(a-c) Mean velocity, $\tu_{00}$. (d-f) Kinetic energy of the retained harmonics.
(a,d) Growth leg. (b,e) Decay. (c,f) Recovery. 
Time increases from red to blue, separated by $t^*=0.025$ among curves. 
}
\label{fig:legbox}
\end{figure}
% ========================================================

This is further analysed in figures \ref{fig:legs}(a-c). Each of these panels represents a
leg of the burst that traverse the periphery of the distribution in the
$(\psi_{v10},I_{v10})$ plane. For example, figure \ref{fig:legs}(a) represents the growing
leg, from the cell marked as L to the one marked as B (note that the latter has been chosen
substantially lower than in figure \ref{fig:orbit}, to bring it within the 95\% probability
contour). It takes the system $T^* \approx 0.25$ to move from one to the other, and the grey
lines are the $O(500)$ trajectories that pass through L at $t=0$. Of these, only the six
trajectories drawn in blue also pass through B in the interval $T^* = 0.2$--0.3. Figures
\ref{fig:legs}(b,c) are similarly drawn for the decay and recovery legs of the burst,
respectively. The trajectories in these three legs are not continuations of each other. For
example, there is a single trajectory linking cells L, B and R, in that order, three
trajectories linking R, L and B, and no trajectory linking B, R and L. The only trajectory
approximating a full cycle in this plane is the one in figure \ref{fig:orbit}.

Figures \ref{fig:legs}(d-f) display the mean evolution of the flow field, reconstructed from
the retained harmonics along the blue trajectories in figures \ref{fig:legs}(a-c),
respectively, with time increasing from top to bottom. Because the problem is homogeneous in
$x$ and $z$, the flow in the different trajectories has to be aligned to a common origin
before averaging. This is done at $t=0$ by translating each case so that the deepest
low-speed perturbation of the streamwise velocity is located at the centre of the display
box. To facilitate visual tracking, the frame of reference is then advected with a velocity
$U_{ad}^+=8$, which keeps the structures approximately stationary near the wall.

The forward tilting of the $v$-structures by the shear is clear in the growth and decay legs
of the burst (figures \ref{fig:legs}d,e). The recovery leg in figure \ref{fig:legs}(f) is
harder to interpret, in part because everything is much weaker than in the other two cases,
but the clearest difference is that, while the growth and decay are dominated by
high-streamwise-velocity structures near the wall, the high speed regions of the recovery
leg are predominantly farther into the flow. The near-wall layer only contains a weak
discontinuous low-speed streak. Since the structures in figure \ref{fig:legs}(d-f) are
defined with respect to a long-term averaged velocity profile, this difference in
organisation implies that the box-averaged instantaneous velocity profile differs among
legs.

This is confirmed in figures \ref{fig:legbox}(a,b), which show that the burst is
characterised by high velocities and steep profiles near the wall. Figure
\ref{fig:legbox}(c) shows that the recovery takes place in regions where the velocity is
lower near the wall, and the shear is displaced away from it. The flow enters the recovery
leg with a steep velocity gradient at $y/h\approx 0.2$, which decays as the recovery
proceeds (red to blue). Figures \ref{fig:legbox}(d-f) show the evolution of the fluctuation
energy at the same times. During the burst, in figures \ref{fig:legbox}(d,e), the
fluctuations stay relatively close to the wall $(y/h\approx 0.12)$, and the energy peak
extends all the way to it. During the recovery leg, in figure \ref{fig:legbox}(f), a new
peak grows at the location of the detached shear layer $(y/h\approx 0.22)$, and forms a new
secondary peak at the wall.

Note that the streamwise velocity perturbations in most frames of figure \ref{fig:legs}(d-f)
are short and discontinuous, in agreement with the recent evidence in \cite{jim22_nostr}
that long streaks are not directly linked to the wall-turbulence regeneration cycle.
 
The conclusion that the recovery of the burst depends on a low-shear region near the wall is
inconsistent with the intuitive notion that, since the shear is the ultimate source of
turbulent energy, a higher shear should be a prerequisite for higher turbulent
activity. Indeed, it has been known for some time that turbulent intensity and shear are
correlated \citep{Mar_Sci10,jim12_arfm,mathis_etal13}, but the implied model is usually
that the turbulence intensity evolves to be in equilibrium with the shear. Our discussion suggests
that the causality is the other way around (figure \ref{fig:condit}), and that the
shear is created by the Reynolds stresses of the fluctuations. In fact, since the mass
flux in our channels is constant, a low shear near the wall implies a higher one further up.
What our previous discussion suggests is that the decay of a burst induces a mild
shear at the wall, which in turn steepens the velocity profile away from it. The steep
off-wall profile is what triggers the new burst, which is the blue/magenta structure growing
away from the wall in the downstream part of figure \ref{fig:legs}(f). \cite{oscar10_log} showed
that stress waves travel to and from the wall in small-box simulations such
as the present one, with a wall-normal velocity of the order of $\utau$. It is difficult to
decide from such kinematic observations which of the two directions is the primary causal
one, but the discussion above suggests that at least the descending wave is causal, in
agreement with previous reports that the structure of the logarithmic layer in wall-bounded
flows is relatively independent from the details of the wall, which is therefore not the
primary, or at least not the only, seat of causality \citep{Townsend1976,miz:jim:13,Kwon21}.
  
% -----------------------------------------------------------------------------
\section{An interventional experiment}\la{sec:meanprof}

The analysis in the previous sections gives strong hints about which variables evolve
coherently in wall-bounded flows, and about how this evolution can be interpreted in terms
of causality within the attractor in phase space. However, the second question posed in the
introduction, whether this information has any practical value, generally takes us outside
the attractor, and can only be answered by more classical interventional experiments or by
theoretical models. Strictly speaking, this is beyond the scope of the present paper, whose
goal is to develop a methodology and to give examples of how it can be used to motivate more
classical work, but we present in this section an example of how that work might proceed,
building on our discussion of figure \ref{fig:legs}.

The main result of that discussion is that a particular, weakly sheared, configuration
of the mean velocity profile is a requirement for burst recovery. This is not
a new idea: what is probably the oldest theory of how wall turbulence is controlled
is based on the two-way interaction between shear and turbulence intensity \citep{malkus56}. It
is also known that forcing a locally steeper or shallower velocity profile leads to equilibrium intensities
that correlate with the local shear \citep{tue:jim:13}, and many low-order models of the 
turbulence cycle include a variable that stands for the mean velocity \citep{Waleffe97}.      

However, the original linear version of \cite{malkus56} model was disproved by
\cite{reytied67}, who found no trace of the marginal instability of the velocity profile
that that model proposes for turbulent channels, and the analysis in \cite{Waleffe97}, although
highly suggestive because the effect of a strong wall shear is to inhibit the instability of
the streaks, only applies to permanent travelling waves in marginally turbulent low-Reynolds
number flows. Similarly, \cite{tue:jim:13} refer to long-term flow averages,
and it is unclear whether these are relevant to the short-term behaviour of an intermittent
bursting cycle. The hypothesis to be tested is whether a feed-back cycle involving the
modification of the mean profile by the turbulent fluctuations, and the control of the
latter by the former, can explain at least part of the mechanism that sets the frequency and
amplitude of the bursts.

% ===========================================================
\begin{figure}
%\vspace*{1mm}%
%\centerline{%
%%
%\raisebox{0mm}{\includegraphics[width=0.48\textwidth,clip]%
%{\figpath cfhisT_0_50.pdf}}%
%\mylab{-.05\textwidth}{.13\textwidth}{(a)}%
%%
%\hspace{2mm}%
%%
%\raisebox{0mm}{\includegraphics[width=0.45\textwidth,clip]%
%{\figpath enerhis5T_0_50.pdf}}%
%\mylab{-.05\textwidth}{.13\textwidth}{(b)}%
%}%%
%%
%\vspace*{2mm}%
%%%%%%%%%%%%%%%%%%%%%%%%%%%
%\centerline{%
%%
%\raisebox{0\textwidth}{\includegraphics[width=.28\textwidth,clip]%
%{\figpath fixprof0T_0_50.pdf}}%
%\mylab{-.21\textwidth}{.24\textwidth}{(c)}%
%%
%\hspace*{12mm}%
%%
%\raisebox{.0\textwidth}{\includegraphics[width=.27\textwidth,clip]%
%{\figpath fixprof13T_0_50.pdf}}%
%\mylab{-.05\textwidth}{.235\textwidth}{(d)}%
%%
%}%%
%%
%\vspace*{5mm}%
%%%%%%%%%%%%%%%%%%%%%%%%%%%
%\centerline{%
%\hspace*{3mm}%
%\raisebox{0\textwidth}{\includegraphics[width=.44\textwidth,clip]%
%{\figpath Uavhis_mini950his.pdf}}%
%\mylab{-.20\textwidth}{.125\textwidth}{(e)}%
%\hspace{5mm}%
%%
%\raisebox{0.005\textwidth}{\includegraphics[width=.48\textwidth,clip]%
%{\figpath Uavhis_mini950fixT50.pdf}}%
%\mylab{-.29\textwidth}{.125\textwidth}{(f)}%
%%
%}
%%
%\vspace{4mm}%
%%
%\centerline{%
%\raisebox{0mm}{\includegraphics[width=.45\textwidth,clip]%
%{\figpath Khis_mini950his.pdf}}%
%\mylab{-.215\textwidth}{.175\textwidth}{(g)}%
%\hspace{3mm}%
%%
%\raisebox{0mm}{\includegraphics[width=0.45\textwidth,clip]%
%{\figpath Khis_mini950fixT50.pdf}}%
%\mylab{-0.26\textwidth}{.175\textwidth}{(h)}%
%%
%}%%
\includegraphics[width=1\textwidth,clip]{\figpath 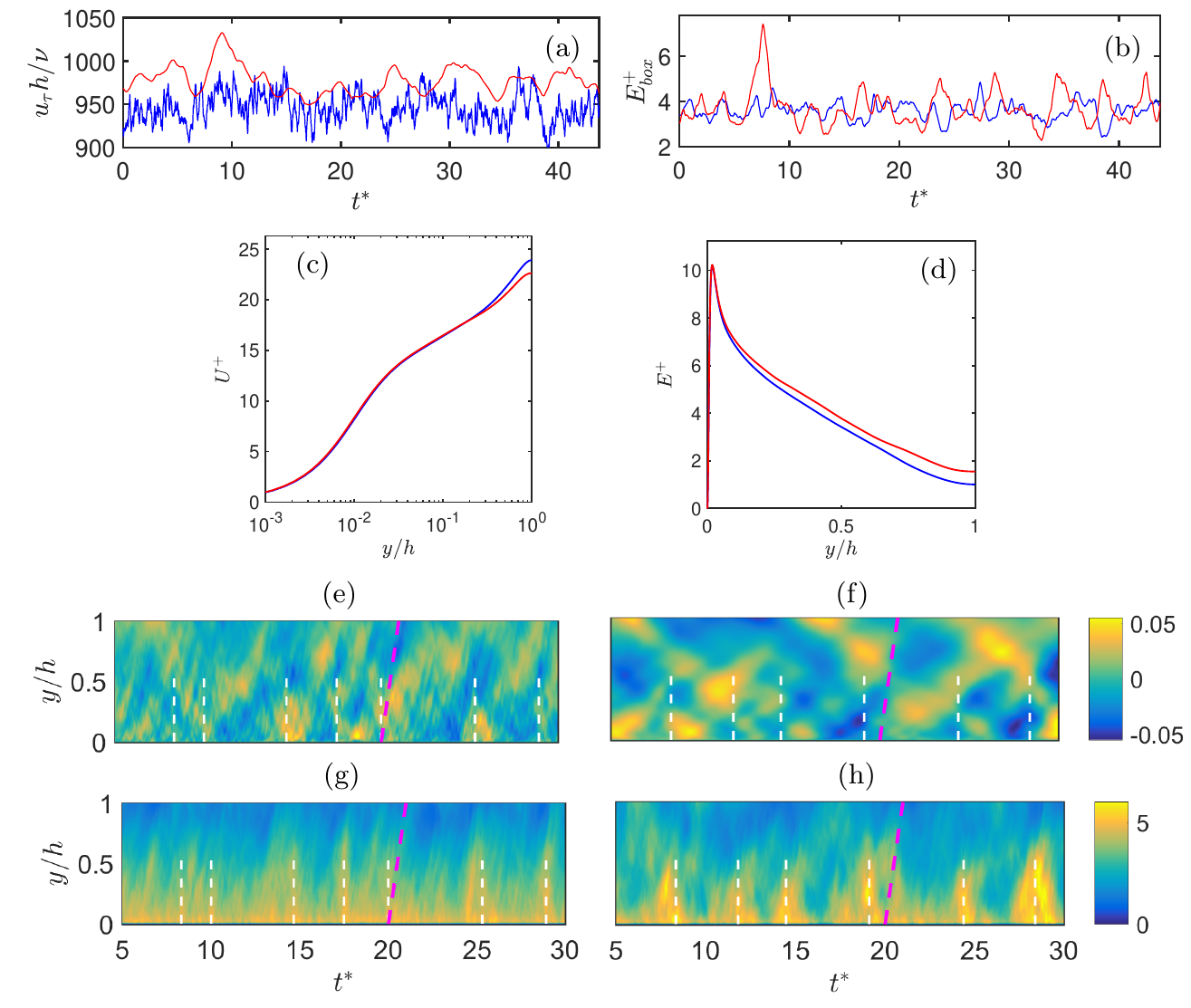}
\caption{%
(a) Instantaneous friction velocity of the temporally smoothed and natural experiments,
(\ref{eq:RQ1}--\ref{eq:RQ4}).
(b) As in (a), for the box-averaged fluctuating kinetic energy, $E=u'^2+v'^2+w'^2$, 
measured with respect to its long-time average. 
(c) Mean velocity profiles.
(d) Mean kinetic energy.
In all cases: blue, natural channel; red, $\tau^*=2.27$.  %; black, $\utau\tau/h=9.3$. 
(e-h) Temporal evolution of the mean profiles as functions of time. 
(e,f) Mean streamwise velocity fluctuation,
with respect to its long-time average. 
(g,h) Kinetic energy. 
(e,g) Natural channel. (f,h) Temporally smoothed one. 
}
\label{fig:fixprofile}
\end{figure}
% ========================================================

We do this by smoothing the evolution equation that links the fluctuations to the mean
profile, defined as the deviation from the long-term mean velocity profile of the
box-averaged streamwise velocity fluctuation, $\tu_{00}(y)$. In a regular channel, it
satisfies,
\beq
\p_t\tu_{00} = R-{P} + \p_{yy} \tu_{00},
\la{eq:RQ1}
\eeq
where $R=-\p_y(\widetilde{uv})_{00}$ is the instantaneous mean Reynolds-stress gradient, and
the pressure gradient $P$ is determined by the ancillary flux-conservation condition,
\beq
\int_0^{2h} \tu_{00} \dd y= 0
\la{eq:RQ2}
\eeq
In our experiment, we substitute \r{eq:RQ1} by
\beq
\p_t\tu_{00} = Q-{P} + \p_{yy} \tu_{00},
\la{eq:RQ3}
\eeq
where
\beq
\p_tQ = (R-Q)/\tau,
\la{eq:RQ4}
\eeq
which, after a transient in which the effect of the initial conditions decays exponentially, is solved by
\beq
Q(y,t) = \tau^{-1} \int_0^t \exp[(\xi-t)/\tau] \, R(y,\xi) \dd\xi.
\la{eq:RQ5}
\eeq
The modified right-hand side, $Q$, is therefore a smoothed version of $R$, with a smoothing
time $\tau$. The integral of $R$ or $Q$ across the channel can be considered as a body force
that must be compensated by the pressure gradient, but it is easy to see from its expression
that $\int R\dd y =0$ for impermeable walls, and that the same holds for $Q$ after the
initial transient.

Figure \ref{fig:fixprofile} shows some results from the experiment. Figure
\ref{fig:fixprofile}(a) shows the history of the friction Reynolds number for the natural
and modified channels. It only increases slightly for the smoothed case, from $\bra
h\ket^+=950$ to 975, although its temporal oscillations become much slower. More interesting
is figure \ref{fig:fixprofile}(b), which shows the oscillations of the box-averaged kinetic
energy. They are somewhat deeper for the smoothed case, and substantially less frequent and
more regular. An approximate count, using a method explained below, gives $\Delta t^*\approx
3.9$ for the mean distance between bursts in the natural case, and $\Delta t^*\approx 5.2$
in the smoothed one. Both are longer than in \cite{oscar10_log}, who find $\Delta t^*\approx
2$ from the temporal spectrum of the integrated Reynolds stress, probably because their method is
sensitive to weaker oscillations than the present one.

The effect on the integrated velocity profiles is slight. Figure \ref{fig:fixprofile}(c)
shows the mean velocity, and reveals that the main effect is to decrease the wake component
above $y/h\approx 0.3$, but this is also the height at which this channel begins to be
constrained by the numerical box, and where the profile in any case deviates from the
natural one. Figure \ref{fig:fixprofile}(d) shows that the fluctuations of the kinetic
energy are also slightly higher, as expected from figure \ref{fig:fixprofile}(b).

Figures \ref{fig:fixprofile}(e-h) show the temporal evolution of the profiles, and give more
information. Figures \ref{fig:fixprofile}(e,f) show the deviation, $\tu_{00}$, of the mean
velocity profile with respect to its long-time average. Each vertical section of these
images is an instantaneous box-averaged profile. Blue regions are slower that usual, and
yellow ones are faster. They should be compared to the instantaneous profiles in figure
\ref{fig:legbox}(a-c), although the flows in the present figure are not filtered or
conditioned in any way. Figures \ref{fig:fixprofile}(e,g) are the natural flow, and figures
\ref{fig:fixprofile}(f,h) are temporally smoothed. The white dashed vertical lines in figure
\ref{fig:fixprofile}(e-h) mark the time of the bursts of the kinetic energy, whose evolution
is represented in figures \ref{fig:fixprofile}(g,h). To detect them, the kinetic energy is
integrated in $y/h\in(0,0.4)$, and bursts are defined as intervals where the integrated
energy rises above the level isolating the top 15\% of the time.
 
% ===========================================================
\begin{figure}
%\vspace*{5mm}%
%%
%\centerline{%
%\begin{minipage}[b]{0.80\textwidth}
%%
%\raisebox{0mm}{\includegraphics[width=1\textwidth,clip]%
%{\figpath TempUav_mini950fixT50.pdf}}%
%\mylab{-1.07\textwidth}{.085\textwidth}{(a)}%
%
%%
%\raisebox{0mm}{\includegraphics[width=.94\textwidth,clip]%
%{\figpath TempK_mini950fixT50.pdf}}%
%\mylab{-1.02\textwidth}{.085\textwidth}{(b)}%
%
%%
%\raisebox{0mm}{\includegraphics[width=0.97\textwidth,clip]%
%{\figpath Tempuv_mini950fixT50.pdf}}%
%\mylab{-1.05\textwidth}{.17\textwidth}{(c)}%
%\end{minipage}%
%}%
\includegraphics[width=1\textwidth,clip]{\figpath 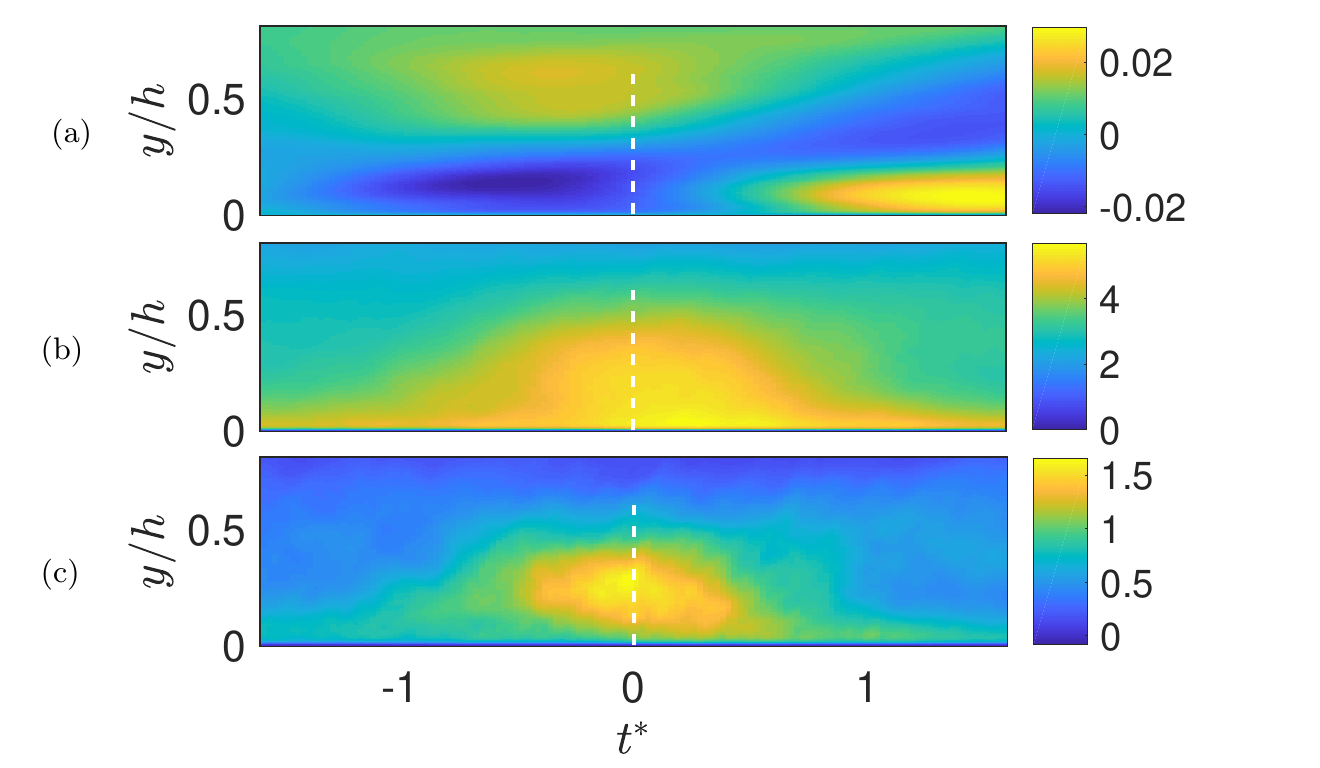}
\caption{%
Time evolution of the box-averaged profiles conditioned to the bursts of the kinetic energy in the
smoothed experiment. 
(a) Mean streamwise velocity fluctuation, as in figure \ref{fig:fixprofile}(f).
(b) Kinetic energy. This is the conditioning field.
(c) Tangential Reynolds stress, $-(\widetilde{uv})_{00}^+$. 
}
\label{fig:fixtemplates}
\end{figure}
% ========================================================

The most interesting differences are those between the mean velocities in figures
\ref{fig:fixprofile}(e, f). The evolution in the natural case in figure
\ref{fig:fixprofile}(e) is clearly more complex than the temporally smoothed case in figure
\ref{fig:fixprofile}(f), and there is no clear correlation between the mean velocity and the
position of the bursts marked by the dashed white lines. The opposite is true for the
smoothed case in figure \ref{fig:fixprofile}(f), in which the mean profile rises and falls
in a series of diagonal waves, and the lines marking the bursts correspond, even to the
naked eye, to low-velocity intervals marked by bluish areas near the wall. The inclined red
line in the four panels of the figure mark the friction velocity, $\dr y/\dr t=u_\tau$,
which is known to be the vertical advection velocity of strong Reynolds-stress structures
\citep{oscar10_log, lozano-time}. This agrees with the inclination of the rising and falling
patterns in the energy evolution maps in figures \ref{fig:fixprofile}(g,h). It also
approximately describes the vertical advection velocity of the fine structure of the
velocity profiles in figure \ref{fig:fixprofile}(e), most probably because the mean profile
is controlled by the Reynolds stress through \r{eq:RQ1}. On the other hand, this influence
is broken in figure \ref{fig:fixprofile}(f), where the Reynolds stresses only acts
indirectly on the mean profile because of the smoothing effect of \r{eq:RQ4}, and the
vertical advection is much slower. This strongly suggests that any correlation between
figures \ref{fig:fixprofile}(f) and \ref{fig:fixprofile}(h) reflects a causal effect of the
mean velocity on the burst, rather than the other way around.

This is directly tested in figure \ref{fig:fixtemplates}, which shows the conditionally
averaged temporal evolution of different quantities obtained by centring them on the
time of the detected energy bursts. Figure \ref{fig:fixtemplates}(a) shows the
conditional evolution of the mean profile, and clearly shows the low-velocity period
preceding the burst, which is later substituted by a steeper wall profile created by the Reynolds
stress of the burst, shown in figure \ref{fig:fixtemplates}(b). The conditional Reynolds
stress is displayed in figure \ref{fig:fixtemplates}(c), which shows that it is a local
effect due to the burst itself. An attempt to repeat this process for the
natural flow in figures \ref{fig:fixtemplates}(e,g) fails, no doubt in part because the
more complex structure of the flow field makes the identification of the bursts harder. In
fact, even the conditioning of the burst on the burst position, as in figure
\ref{fig:fixtemplates}(b), fails in that case.

% -----------------------------------------------------------------------------
\section{Conclusions}\la{sec:conc}

This paper can be divided in two parts. In the first one, up to \S \ref{sec:data}, we adapt
the Perron--Frobenius operator (PFO) of dynamical-system theory to the probabilistic
description of the evolution over a phase space partition of a turbulent channel flow. We
show that the main difficulty for doing so is collecting enough data to populate the
operator matrix, and we bypass it by restricting ourselves to two-dimensional projections of
the phase space. This leads to the question of how to choose the best pair of variables, and
forces us to develop simple indicators of the quality of a particular representation.
Several such indicators are developed, and shown to be interpretable in terms of causality
and coherence within the attractor. It is argued that this last restriction allows us to
draw conclusions about causality and information flux from flow histories, without the need
for interventional experiments.

In particular, we show that we can use these indicators to distinguish between correlation
and coherence, and to separate, for example, relatively weak structures that have their own
dynamics, such as wavy rollers and streaks, from stronger and more correlated ones
that have no dynamics of their own, such as streamwise-uniform streaks and rollers.

We show in \S \ref{sec:drift} how the indicators allow us to differentiate less promising
variable pairs from those more likely to be useful in developing coherent physical models.
Out of 630 possibilities, two promising pairs are found for the case analysed here. The
first one is the intensity and inclination of the wall-normal velocity, which was already
used by \cite{jimenez:2015} to represent an approximately linear \cite{orr07a} burst, and
the second is a more novel inclined wavy vortex.

The rest of the paper applies the techniques derived in the first part to analyse the
\cite{orr07a} burst, with emphasis on the poorly understood recovery process by which bursts
are re-initiated after they decay. As was the case with previous attempts to use massive
searches to choose among different analysis possibilities
\citep[e.g.,][]{jimploff18,jotploff}, the present one mostly suggests mechanisms that have
to be confirmed by more classical means, mainly because of the original limitation to
on-attractor dynamics. In this case, the PFO guides us in the choice of phase-space
trajectories that connect interesting flow configurations within a known range of time
intervals, including trajectories describing the recovery process. At least in our
relatively small computational box, conditional averaging over these connections shows,
somewhat counter-intuitively, that the key ingredient for regeneration is the development of
a low-shear region near the wall. New bursts are seeded from a detached shear layer
overlying it. Their Reynolds stress returns the shear to the wall, and no new burst is
possible until the decay of the old one again detaches the shear. It is not known whether this
process generalises to larger boxes containing more than one burst.

To extend our conclusions outside the attractor, we finally perform a simple computational
experiment in which the control of the mean shear by the burst is relaxed. The behaviour of
the mean profile is thus modified, but the association of low wall shear with the initiation
of the bursts is shown to be maintained.

%%%%%%%%%%%%%%%%%%%%%%%%%%%%%%%%%%%%%%%
\vspace{2ex}
This work was supported by the European Research Council under the Caust grant
ERC-AdG-101018287. 
%
%\newpage
% ----------------------------------------------------------------------------------------------------
%\bibliographystyle{jfm}
%\bibliography{cit}                          

\appendix
    
% ------------------------------------------------------------
\section{Modes with non-zero spanwise wavenumber}\la{sec:11modes}

The modes with non-zero spanwise wavenumber require special treatment. For $k_z=0$, the
variable $a$ is expressed as
\beq
a(x)=\sum_\alpha \ta_\alpha \ee^{\ii \alpha x} + \mbox{c.c.},
\la{eq:11mode1}
\eeq
where ``c.c.'' stands for the complex conjugate. A translation $x\to x+\Delta x$ appears as
a phase in $\ta_\alpha \to \ta_\alpha \exp(\ii\alpha \Delta x)$, and the variables used in
the main text are chosen to be invariant to such translations. For modes depending on $z$,
the expansion becomes,
\beq
a(x,z)=\sum_{\alpha,\beta} \ta_{\alpha,+\beta} \ee^{\ii (\alpha x+\beta z)} + 
      \ta_{\alpha,-\beta} \ee^{\ii (\alpha x-\beta z)} +
\mbox{c.c.}.
\la{eq:11mode2}
\eeq
Each $(\alpha,\beta)$ mode has associated two independent coefficients, which transform
differently under spanwise translations. The physically relevant expansion is 
\beq
a(x,z)=\sqrt{2}\,\sum_{\alpha,\beta} \left[\ta^+_{\alpha,\beta}\cos(\beta z) + 
      \ii \ta^-_{\alpha,\beta}\sin(\beta z)\right ] \ee^{\ii \alpha x}  +
\mbox{c.c.},
\la{eq:11mode3}
\eeq
where $\ta^\pm_{\alpha,\beta}=\ta_{\alpha,+\beta}\pm \ta_{\alpha,-\beta}$. We can choose to
observe the flow at any convenient spanwise location. Keeping the cosine term in
\r{eq:11mode3} corresponds to observing at $z=0$, while keeping the sine corresponds to
$\beta z =\pi/2$, but neither $\ta_{\alpha,+\beta}$ nor $\ta_{\alpha,-\beta}$
are physical observations. The new coefficients, $\ha^\pm_{\alpha,\beta}$, are also
independent Fourier coefficients, and are used throughout the manuscript. 

% ------------------------------------------------------------
\section{Proper orthogonal filtering}\la{sec:PODs}

% ===========================================================
\begin{figure}
%\vspace*{5mm}%
%\centerline{%
%%
%\raisebox{0mm}{\includegraphics[height=.30\textwidth,clip]%
%{\figpath fracPOD.pdf}}%
%\mylab{-.05\textwidth}{.23\textwidth}{(a)}%
%\hspace*{1mm}%
%%
%\raisebox{0mm}{\includegraphics[height=.30\textwidth,clip]%
%{\figpath upods_1_5_20.pdf}}%
%\mylab{-.05\textwidth}{.23\textwidth}{(b)}%
%\hspace*{1mm}%
%%
%\raisebox{0mm}{\includegraphics[height=.30\textwidth,clip]%
%{\figpath wpods_1_5_20.pdf}}%
%\mylab{-.05\textwidth}{.23\textwidth}{(c)}%
%}%
%%
%\vspace{2mm}%
%%
%\centerline{%
%%
%\raisebox{0mm}{\includegraphics[height=.30\textwidth,clip]%
%{\figpath uprofpod_1_5_20.pdf}}%
%\mylab{-.05\textwidth}{.26\textwidth}{(d)}%
%\hspace*{1mm}%
%%
%\raisebox{0mm}{\includegraphics[height=.30\textwidth,clip]%
%{\figpath vprofpod_1_5_20.pdf}}%
%\mylab{-.05\textwidth}{.26\textwidth}{(e)}%
%\hspace*{1mm}%
%%
%\raisebox{0mm}{\includegraphics[height=.30\textwidth,clip]%
%{\figpath wprofpod_1_5_20.pdf}}%
%\mylab{-.05\textwidth}{.26\textwidth}{(f)}%
%}%
\includegraphics[width=1\textwidth,clip]{\figpath 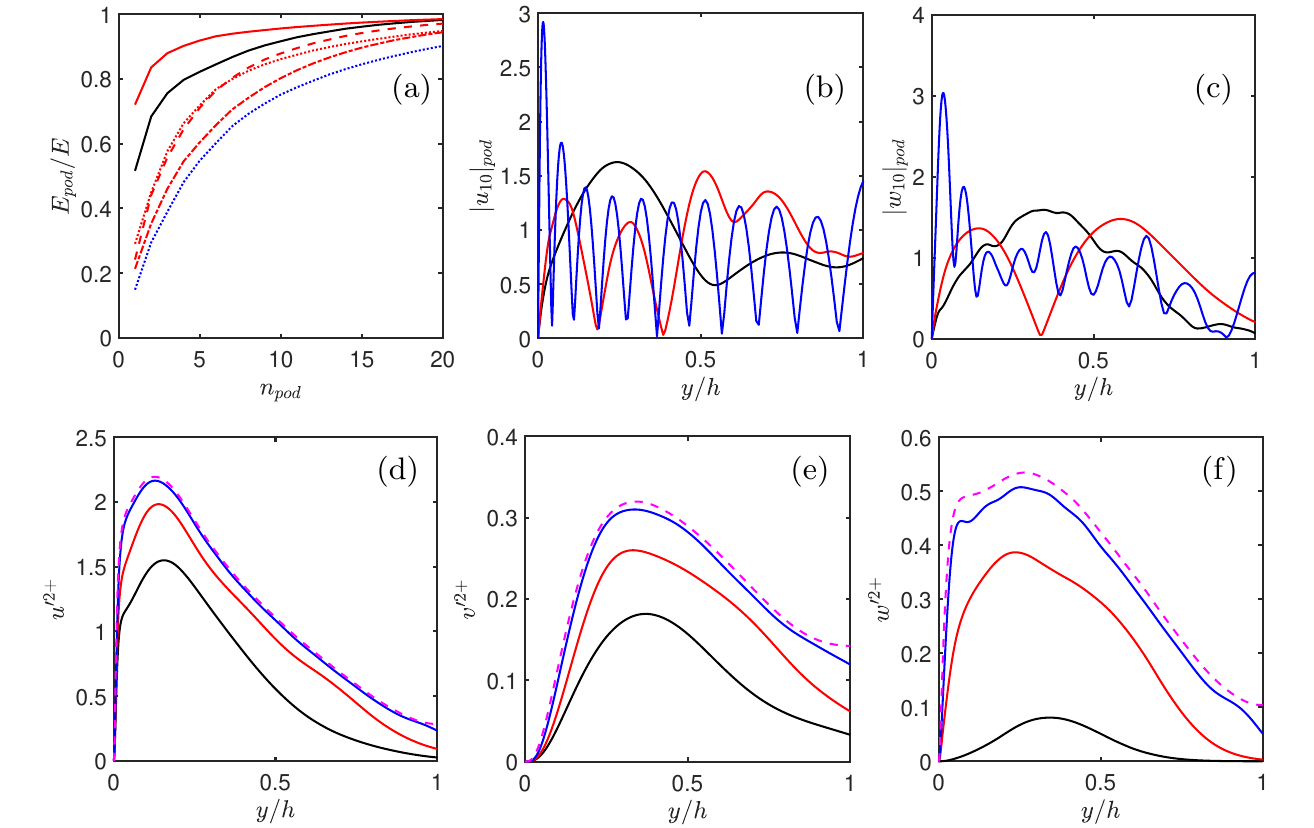}
\caption{%
(a) Energy fraction of the different harmonics contained in first $n_{pod}$ modes. Solid
black, [00] mode; solid red, [10]; dashed red, [10]; chaindotted red, [20]; dotted red,
[11]; dotted blue, [21].
(b, c) Wall-normal intensity profiles of typical POD modes of the [10] harmonic. Black,
$n_{pod}=1$; red, $n_{pod}=5$; blue, $n_{pod}=20$. (b) $u_{10}$. (c) $w_{01}$.
(d-f) Intensity profiles of all the retained harmonics, approximated with different number of
POD modes. Colours as in (b). The dashed magenta line, is the true intensity profile. 
(d) $u$. (e) $v$. (f) $w$.
}
\label{fig:podener}
\end{figure}
% ========================================================

% ===========================================================
\begin{figure}
%\vspace*{5mm}%
%\centerline{%
%%
%\raisebox{0mm}{\includegraphics[height=.30\textwidth,clip]%
%{\figpath wxy_0vp4_va4_t6_6_15X13_nP20_0_14_2_4_2_ig7609.png}}%
%\mylab{-.27\textwidth}{.31\textwidth}{(a)}%
%\hspace*{1mm}%
%%
%\raisebox{0mm}{\includegraphics[height=.30\textwidth,clip]%
%{\figpath wxy_0vp4_va4_t6_6_15X13_nP20_20_14_2_4_2_ig7609.png}}%
%\mylab{-.27\textwidth}{.31\textwidth}{(b)}%
%}%
\includegraphics[width=1\textwidth,clip]{\figpath 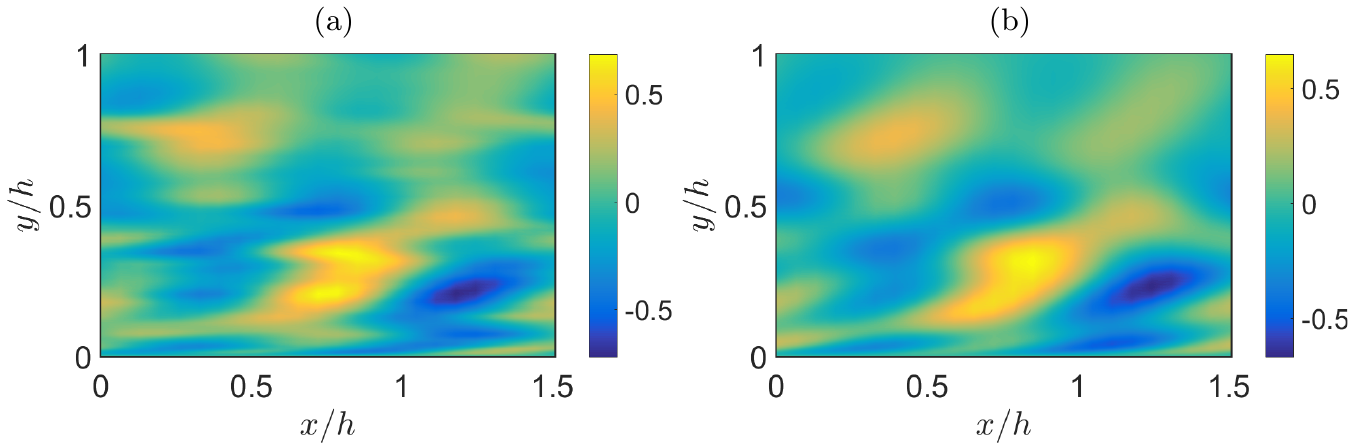}
\caption{%
Example of the effect of POD filtering on a streamwise section of the $w$ field.
(a) Reconstructed with the nine retained Fourier modes and full resolution in $y$.
(b) Using twenty PODs in $y$.
}
\label{fig:podfield}
\end{figure}
% ========================================================

The proper orthogonal decomposition (POD) was introduced to fluid mechanics by
\cite{Ber:Hol:Lum:93} as a variant of the older principal component analysis approximation
of \cite{pearson}. It seeks to represent a vector field $\uvec$ as an expansion of POD modes
$\phivec_{(k)}$,
\beq
\uvec_{(n)} = \sum_{k=1}^n \hu_k \phivec_{(k)},  
\la{eq:POD1}
\eeq
in such a way that the two-point covariance of $\uvec_{(n)}$ approximates as well as
possible the covariance of the true vector field, even when the order of the expansion is
much less than the number of degrees of freedom in $\uvec$. As already mentioned in \S \ref{sec:PF}
and \S \ref{sec:data}, POD modes are not good choices to reduce a dynamical system to a
few variables, because they minimise the intrgral of the error over the attractor as a whole, instead
of, as has been our interest in this paper, differentiating among different phase-space
neighbourhoods. However, they can approximate the flow field using a somewhat
reduced number of spatial modes.

It is clear that our choice to represent the flow using nine Fourier modes in the $(x,z)$
plane is a drastic reduction of numerical resolution in that plane, but there is no simple
equivalent way of reducing the resolution along the non-homogeneous direction, $y$. The
result is a strongly anisotropic flow field with spurious thin wall-parallel layers which
are not justified by the wall-parallel resolution. POD modes provide a useful basis to
balance the resolution in a statistically significant way. In fact, Fourier modes are POD
modes along homogeneous directions, such as $x$ and $z$, and we saw in figure \ref{fig:uprof} that the few retained modes
account for a significant fraction of the kinetic energy \citep[most of the mathematical
properties mentioned in this appendix are drawn from][although this will not be mentioned from
now on]{Ber:Hol:Lum:93}. The POD analysis can be extended
to $y$ by expanding individual Fourier modes, and figure \ref{fig:podener}(a) shows the
fraction of the total energy of each Fourier mode contained in its first $n$ POD modes. The
first few modes do not represent the energy well, but approximately twenty modes account for
most of it. Because the POD modes in \r{eq:POD1} are eigenvectors of the covariance matrix,
successive eigenvectors contain an increasing number of `waves', as shown in the examples in
figure \ref{fig:podener}(b,c). Broadly speaking, the twenty modes mentioned above are
equivalent to a wall-normal resolution of ten points, which is a reasonable compromise for
the number of wall-parallel Fourier modes. Figure \ref{fig:podener}(d-f) show that this
resolution is enough to account for most of the energy profile of the retained Fourier
approximation.
  
Several precautions are necessary for a consistent approximation. In the first place,
expansion \r{eq:POD1} optimises the approximation of the flow field in terms of the $L_2$
norm defined over the collocation nodes of the numerical grid, while it makes more physical
sense to use an integral norm $\int |\uvec|^2 \dd y$. This requires scaling the flow field
by the square root of the grid spacing, and undoing the scaling upon reconstruction.
Similarly, although figure \ref{fig:podener} displays modes as associated to individual
velocity components, it is important to compute the PODs over the three velocities at the
same time, catenated as a single compound vector. This minimises the total kinetic energy of
the approximation error, and ensures, among other things, that the reconstructed field
satisfies continuity.

An example of the effect of filtering the $w$ field with the first twenty PODs is shown in
figure \ref{fig:podfield}. Most of the results in this paper are obtained from flow fields
filtered with twenty POD modes. There is very little difference between these results and
those from unfiltered variables, except for the somewhat cleaner flow fields such as the one
in figure \ref{fig:podfield}(b).

\end{document}